\documentstyle[12pt]{report}

\def\lromn#1{\uppercase\expandafter{\romannumeral#1}}

\def\blist{\begin{list}{\setlength{\rightmargin}{\leftmargin}}}
\def\elist{\end{list}}

\def\thebibliography#1{\list
 {[\arabic{enumi}]}{\settowidth\labelwidth{[#1]}\leftmargin\labelwidth
 \advance\leftmargin\labelsep
 \usecounter{enumi}}
 \def\newblock{\hskip .11em plus .33em minus .07em}
 \sloppy\clubpenalty4000\widowpenalty4000
 \sfcode`\.=1000\relax}
\let\endthebibliography=\endlist
\def\AP{{\sl Ann.\ Phys.\ {\rm(}N.Y.{\rm)} }}
\def\CMP{{\sl Commun.\ Math.\ Phys.\ }}
\def\FP{{\sl Fortsch.\ Phys.\ }}
\def\GRG{{\sl Gen.\ Rel.\ Grav.\ }}
\def\JMP{{\sl J.\ Math.\ Phys.\ }}
\def\JPSJ{{\sl J.\ Phys.\ Soc.\ Jpn.\ }}
\def\LNC{{\sl Lett.\ Nuovo Cim.\ }}
\def\LNCI{{\sl Lett.\ Nuovo Cim.\ }(Ser.~I), }
\def\MPL{{\sl Mod.\ Phys.\ Lett.\ }}
\def\MPLA{{\sl Mod.\ Phys.\ Lett.\ A }}
\def\NC{{\sl Nuovo Cimento }}
\def\NCA{{\sl Nuovo Cimento A }}
\def\NP{{\sl Nucl.\ Phys.\ }}
\def\NPB{{\sl Nucl.\ Phys.\ B }}
\def\PL{{\sl Phys.\ Lett.\ }}
\def\PLA{{\sl Phys.\ Lett.\ A }}
\def\PLB{{\sl Phys.\ Lett.\ B }}
\def\PRep{{\sl Phys.\ Rep.\ }} \let\PREP=\PRep
\def\PR{{\sl Phys.\ Rev. }}
\def\PRA{{\sl Phys.\ Rev.\ A }}
\def\PRD{{\sl Phys.\ Rev.\ D }}
\def\PRL{{\sl Phys.\ Rev.\ Lett.\ }}
\def\PS{{\sl Physica Scripta }}
\def\PTP{{\sl Prog.\ Theor.\ Phys.\ }}
\def\PTPS{{\sl Prog.\ Theor.\ Phys.\ Suppl.\ }}
\def\ZPC{{\sl Z.\ Phys.\ C\ }}
\begin{document}

\begin{titlepage}

\begin{flushright}
\begin{large}
TU/94/452\\
\end{large}
\end{flushright}

\vspace{1.5cm}

\begin{center}
\begin{Large}

\renewcommand{\thefootnote}{\fnsymbol{footnote}}
\bf{Moving Mirror Model of Hawking Evaporation}
\footnote[2]
{Submitted for publication to Progress of Theoretical Physics}

\end{Large}

\vspace{1cm}

\begin{large}
\renewcommand{\thefootnote}{\fnsymbol{footnote}}
M.Hotta, M.Shino,
and M.Yoshimura\footnote[3]
{E-mail address: YOSHIM@tuhep.phys.tohoku.ac.jp
}\\
Department of Physics, Tohoku University\\
Sendai 980 Japan\\

\vspace{3.5cm}

{\bf Abstract}\\

\end{large}
\end{center}

\vspace{0.5cm}

The moving mirror model is designed to extract essential features of
the black hole formation and the subsequent Hawking radiation by neglecting
complication due to a finite curvature. We extend this approach
to dynamically treat
back reaction against the mirror motion due to Hawking radiation.
It is found that a unique model in two spacetime dimensions exists
in which Hawking radiation completely stops and the end point of evaporation
contains a disconnected remnant.
When viewed from asymptotic observers at one side of the spacetime,
quantum mechanical correlation is recovered in the end.
Although the thermal stage accompanying short range correlation
may last for an arbitrarily long period,
at a much longer time scale a long tail of
non-thermal correlation is clearly detected.

\vspace{12pt}

\end{titlepage}

\addtocounter{chapter}{1}
\setcounter{equation}{0}
\section*{\lromn
1. Introduction
}

\vspace{0.5cm}
\hspace*{0.5cm}
The end point of Hawking evaporation process poses a challenging problem
of fundamental nature; a purely quantum mechanical massive state can
gravitationally collapse, forming the event horizon, from which apparently
thermal emission of radiation follows \cite{hw75} .
Does this process continue until
complete evaporation? If this is the inevitable outcome, a pure quantum
state can evolve into a mixed state and the fundamental
premise of quantum mechanics is clearly violated \cite{hw76},
but how is it violated? On the other hand, if this is not the case,
how is the Hawking evaporation terminated and
how is the lost information during the almost thermal stage recovered?

Until recently, this important problem has not been analyzed in sufficient
detail to unambiguously resolve the issue.
One of the main reasons for the lack of convincing arguments is
that the important element in the analysis,
before one makes a definite conclusion, is still missing: the back reaction
problem. It is a formidable task, in any realistic situation, to incorporate,
beyond the leading semiclassical approximation,
the back reaction against the spacetime structure due to Hawking radiation.
Unless one can follow the back reaction till the end point, it is however
difficult to accept without reservation any proposal for the resolution.

Recent upsurge of interest in this problem stems from the observation
that the essential features of the spherical collapse may be retained in
a class of two dimensional dilaton gravity theories \cite{cghs}
in which the back
reaction may be analyzed in considerable detail. Yet no consensus on the
end point behavior emerged despite many interesting works \cite{rstetc}.

In this paper we shall adopt the moving mirror picture \cite{witt75}
to further advance
the analysis of the end point behavior of Hawking evaporation.
It is well known \cite{bdavies}, \cite{wilcz93} that the
accelerating mirror can induce the Hawking radiation from vacuum in
essentially the same manner as in the black hole case. One can sharpen
this correspondence in the case of dust shell collapse \cite{unruh},
\cite{wilcz93}.
Thus at least in the semiclassical level the moving mirror model is an
excellent approximation to the Hawking radiation. The remaining task in this
picture is to dynamically treat the back reaction against the mirror particle
due to Hawking radiation
and is to remedy the shortcoming of the mirror approach, namely
to include the neglected effect of curvature.
In this paper we shall take up
one of these tasks, and incorporate
the back reaction by exploiting experience gained
by two dimensional toy models. We shall mainly examine a particular
underlying dilaton model, although our treatment of the back reaction may
be applied to other similar models, and indeed we have some results
in a variety of other models.

We shall state our main physical results at this introductory stage.
Our classical dilaton model \cite{hy93-2} describes the gravitational
collapse of a massive
body, which results in formation of event horizon. This model does not possess
the curvature singularity in the usual sense, although idealization of
the massive source by a point particle does contain a weaker delta-function
singularity. The weak curvature singularity in this case
is presumably an artifact of the
localization, and is expected to be smeared out for an extended source.
Existence of the event horizon gives rise to Hawking radiation, as verified
either by the technique of Bogoliubov transformation \cite{hw75}
or by a calculation of one loop quantum effect  \cite{cf}, \cite{cghs}.

In the corresponding moving mirror picture of this model
the mirror trajectory approaches a null asymptote  at an exponentially rapid
rate, receding from asymptotic observers. This exponential recession
is crucial to derive the thermal spectrum.
In the mirror picture one imposes the reflection boundary condition to
incident waves in order to mimic the
spherically symmetric collapse in which the radial coordinate is restricted
to a half of the infinite one dimensional space.
The canonical quantization of the massless
field having the boundary of the reflecting mirror in two spacetime dimensions
induces the Hawking radiation,
when applied to the mirror trajectory corresponding to formation of the
event horizon.
The thermal particle production is caused by mixing of the positive
and the negative energy states occuring as a consequence of that incoming
waves at very late times are never reflected from the mirror in this case.
This interpretation of Hawking radiation is physically transparant and
well documented in the literature \cite{bdavies}.
Fortunately one can do
much more than this in the mirror picture by closely examining the behavior of
quantum field with the reflection boundary. This is what we wish to mainly
investigate in this work.

Thus the obvious important step is to incorporate the back reaction
against the mirror motion due to the particle
production. It is important to formulate the problem such that
effect of particle production or more generally quantum effect
with the reflection boundary can be taken into account without
assuming a specific mirror trajectory.
We first unambiguously compute the local stress tensor giving rise
to an asymptotic flux produced quantum mechanically and observable
to those sitting far away from the mirror.
We then proceed to determine the force term that expresses in the
mirror equation the back reaction
against the mirror particle by imposing invariance under
reparametrization of the mirror trajectory as in the ordinary point
particle case.
There is some degree of uncertainty in this procedure,
and even in our limited approach there is one undetermined parameter.
We shall investigate how final
physical results depend on this parameter. It turns out that only a
specific choice of this parameter yields an interesting model of
complete Hawking evaporation consistent with quantum mechanical
evolution seen by asymptotic observers.

Once the classical mirror dynamics
is specified by the underlying gravity model such as a two dimensional
truncation of the spherically symmetric collapse or a dilaton model, then
one may follow the fate of mirror particle including the quantum back
reaction.
Under the usual setting of a very massive
collapse this treatment gives rise to Hawking radiation for a long period of
time, but one may further detect decay of the thermal correlation,
depending of course on the underlying gravity model, or
equivalently on the corresponding classical mirror trajectory.
The most urgent issue to settle
is of course the end point behavior of the mirror motion,
from which one can deduce the end point behavior of spacetime and how
quantum mechanical correlation behaves. All this sort of questions may be
answered in our approach by analyzing a single nonlinear differential equation
for the mirror motion
which includes the quantum back reaction.

The specific model we most closely examined predicts a unitary quantum
mechanical development that may be seen to asymptotic observers.
In the mirror picture the trajectory approaches in the end a
null line with a uniform acceleration, in contrast to the growing acceleration
in the case of the event horizon formation. Although there remains
a disconnected
area resembling some kind of a remnant, the outside world recovers
the usual correlation as in the case of a fixed boundary,
and the Hawking radiation stops.
The thermal stage of Hawking radiation accompanying short range correlation
may last for an arbitrarily
long period of time, but there exists a long tail of
non-thermal correlation viewed much beyond this time scale.
Thus the system we have here is a very unusual one quantum mechanically,
being characterized by a long range correlation in spite of the local
thermal behavior.
In all other models we investigated, the unitary evolution breaks
down to the asymptotic observer.

During the course of this work we came across a paper by Chung and
H.Verlinde \cite{verlin93}.
Their work differs from ours, in two technical points and one fundamental
point: technically, first on
how to treat the back reaction and second on the underlying
classical model itself. At the more fundamental level, Chung and Verlinde
extract the mirror model from the two dimensional dilaton theory by
identifying a constant line of the dilaton factor $e^{-2\varphi }$
as the location of the mirror point. This procedure is by no means
unique, and we shall take a somewhat different path.
We shall make a few comments on their work in appropriate places, but let us
state at the very beginning that their final picture is entirely
different from ours; their model fails to terminate Hawking radiation.
Presumably the Planck scale physics is important to settle the issue of
the information loss paradox in their model.

We do not claim that our end point behavior is unique,
but at the very least we may point out that an
interesting toy model of complete Hawking evaporation has been found,
consistent with quantum mechanics, but at the price of remnants.
How we may extend our analysis to more realistic four dimensional cases
is another story, and much left to a further development in the future.

The remainder of this paper is organized as follows.
In Section \lromn2 the moving mirror model is formulated assuming that the
corresponding gravity model is given. In the situation in which the classical
spacetime allows formation of the event horizon it is shown that the mirror
trajectory approaches a null line at an exponential rate. A specific
example of a two dimensional dilaton model is used as an illustration.
Behavior of the quantum field in a curved spacetime is closely related
to the quantum field with a moving boundary.
In Section \lromn3 this subject is reviewed in two spacetime
dimensions to the extent relevant to our later discussion in order to make
this paper self-contained.
How the moving mirror picture describes the essential features of
Hawking radiation is explained in considerable detail.
Also is emphasized a particularly illuminating explanation of
Hawking radiation from a hole theory viewpoint. To determine whether or
not particle production occurs is thus almost equivalent to whether mixing
between the positive and the negative energy states takes place.
In Section \lromn4 we formulate the problem of incorporating the back
reaction against the mirror motion due to particle production
in such a way that the mirror equation of motion is modified
by a local term. In Section \lromn5 the classical part of the mirror
action is given for the underlying two dimensional dilaton model
of our interest, and
the back reaction problem is analyzed and solved for this model.
Both analytical
and numerical methods give a unique end point behavior, as explained
in the preceding paragraphs. Section \lromn6 is devoted to extension
of models. Some of the detailed formulas not crucial to understanding
the main body of arguments is relegated to Appendix A-D.

\addtocounter{chapter}{1}
\setcounter{equation}{0}
\section*{\lromn
2.
Gravitational collapse and moving mirror
}

\vspace{0.5cm}
\hspace*{0.5cm}
The moving mirror model \cite{witt75}, \cite{bdavies},
\cite{wilcz93}, \cite{carlitz}
is introduced to extract essential features of
any realistic gravitational collapse,
in particular the four dimensional spherically
symmetric collapse, discarding inessential complexity of a real situation.
Consider a mirror point moving to the left with a constant velocity in $1+1$
dimensional spacetime. Light rays incident on the mirror from the right
get red shifted after reflection, in much the same
way as rays emerging from a gravitational potential. Imagine next
that the mirror is further accelerated to the left towards the light velocity.
Left moving rays that barely catch up the mirror get reflected,
but are much more red shifted as they hit the mirror more lately.
In the end some rays never reach the null asymptote of the mirror trajectory
as in Fig. 1.
These rays remain left moving forever, and in the corresponding collapse
situation they may be regarded as
permanently trapped within the black hole, or more precisely within the
event horizon of the collapsed object.

In order to precisely formulate the gravitational collapse in the mirror
framework, one must identify
the center of the collapse as the location of the moving mirror.
In the original coordinate of the collapse geometry the center location is
taken to be at a fixed $x^{1} = 0$, or equivalently at
\( \:
x^{+} = x^{-}
\: \) in the light cone coordinate. The region left to the center,
\( \:
x^{+} < x^{-}
\: \), does not exist, since waves are reflected at the mirror.
We have in mind the spherically symmetric collapse in four dimensions
with $x^{1}$ replacing the radial coordinate.
The original collapse spacetime described by $x^{\pm }$ is mapped by a
coordinate transformation
to the mirror spacetime which is assumed to be a Minkowski spacetime.
Since the coordinate transformation cannot generate a nonvanishing
scalar curvature, in this formulation one deals with a flat spacetime alone,
except possibly being curved at the center which might correspond to the
usual curvature singularity.
Note, however, that one can deal with the spacetime with event horizon,
whose presence is the key to Hawking radiation.
The mirror approximation thus ignores a hopefully minor effect,
the curvature  effect.

Using as the mirror Minkowski coordinates the light cone variables
\( \:
\sigma^{\pm } = \sigma^{0} \pm \sigma^{1}
\: \),
one defines the mirror trajectory by
\begin{equation}
\sigma^{+} = p(\sigma^{-}) \,,
\end{equation}
such that the mapping to the collapse spacetime is given by
\( \:
x^{+} = \sigma^{+}
\: \) and
\( \:
x^{-} = p(\sigma^{-})
\: \).
One may take by convention that the past null infinity of the mirror spacetime
is mapped to the past null infinity of the collapse spacetime;
\( \:
p(-\,\infty ) = -\,\infty
\: \).
By this construction the original spacetime metric is given by
\begin{equation}
ds^{2} = -\,d\sigma^{+}d\sigma^{-} = -\,\frac{dx^{+}dx^{-}}{p'(\sigma^{-})} \,,
\end{equation}
with
\( \:
p'(\sigma^{-}) = \frac{d}{d\sigma^{-}}\,p(\sigma^{-})
\:. \)
To ensure the asymptotic flatness,
\( \:
ds^{2} \rightarrow -\,dx^{+}dx^{-}
\: \),
one must take the Minkowski spacetime
locally at the past null infinity as
\( \:
x^{-} \rightarrow  -\,\infty
\: \), which is then equivalent to that \( \:
x^{-} = p(\sigma^{-}) \rightarrow  \sigma^{-}
\: \) as \( \:
\sigma^{-} \rightarrow  -\,\infty
\: \). In the mirror picture this means that we only consider the mirror point
initially at rest;
\( \:
\sigma^{+} = p(\sigma^{-}) \rightarrow \sigma^{-}
\: \), or
\( \:
\sigma^{1} \, \rightarrow
\: \) a constant as \( \:
\sigma^{0}  \, \rightarrow \, -\,\infty
\: \).

The crucial feature for the
presence of the event horizon is existence of a simple zero of $p'$
at a finite $x^{-}$;
\( \:
p' \sim \kappa (x_{H} - x^{-})
\: \)
with $\kappa $ a positive constant.
At the position of the zero $x_{H}$, the flat Minkowski
coordinate $\sigma^{-}$ then behaves like
\begin{equation}
\sigma^{-} \sim  -\,\frac{1}{\kappa }\ln \, (x_{H} - x^{-}) \,,
\end{equation}
since
\( \:
\frac{dx^{-}}{d\sigma^{-}} = p'(\sigma^{-})
\: \).
Thus to an asymptotic observer to whom the relevant flat coordinate
is $\sigma^{\pm }$ it takes an infinite amount of $\sigma^{-}$
time to reach the event horizon at
\( \:
x^{-} = x_{H}
\: \): the characteristic feature of the event horizon.
Furthermore this implies that in the mirror picture the mirror
trajectory asymptotes to a null constant $\sigma^{+}$ line at an exponentially
rapid rate
\( \:
e^{-\,\kappa \sigma^{-}}
\: \), namely,
\begin{equation}
\sigma^{+} = p(\sigma^{-}) \rightarrow
x_{H} - \xi \,e^{-\,\kappa \sigma^{-}} \,,
\label{ehorizon}
\end{equation}
with $\xi >0$.
This and only this feature of the mirror trajectory
is important to derive the thermal
spectrum of Hawking radiation. For instance, an approach to a null line
with a power rate does not correspond to presence of the event horizon in
the collapse geometry.

Although the detailed form of the mirror trajectory
\( \:
p(\sigma^{-})
\: \) is not important in most of our discussion,
it would be useful to mention at this point
that the metric given by
\begin{equation}
ds^{2} = -\,\frac{dx^{+}dx^{-}}{1-\frac{M}{2\lambda}\,e^{2\lambda x^{-}}} \,,
\end{equation}
with \( \:
x^{+} > x^{-}
\: \), is an exact classical solution of a variant of two dimensional
dilaton gravity \cite{hy93-2}.
The parameter $M$ represents the mass of the collapsing body,
while the $\lambda^{2} $ is the fundamental parameter of the
underlying dilaton theory of dimension [mass]$^{2}$,
the cosmological constant.
Actually, this spacetime is only a half of the wormhole spacetime considered
in Ref.\cite{hy93-2}.
The disconnected other half, $x^{1} < 0$, is replaced here by the
reflecting mirror keeping in mind the four dimensional spherically
symmetric collapse.

The event horizon is located at
\( \:
x^{-} = x_{H}
\: \) with
\begin{equation}
x_H = -\,\frac{1}{2\lambda} \ln\frac{M}{2\lambda}\,,
\end{equation}
in this model. The mirror coordinate \( \:
\sigma^{-}
\: \)
is derived by integrating
\begin{equation}
\frac{dx^{-}}{d\sigma^{-}} = 1-\frac{M}{2\lambda}\,e^{2\lambda x^{-}}
\,,
\end{equation}
with the initial condition,
\( \:
x^{-} = \sigma^{-}
\: \) at $\sigma^{-} = -\,\infty $.
Thus the precise relation between the collapse spacetime and
the mirror coordinate is as follows;
\begin{eqnarray}
x^+ = \sigma^+ \,, \hspace{0.5cm}  x^- = p(\sigma^{-}) =
-\,\frac{1}{2\lambda} \ln\,(\,\frac{M}{2\lambda} +e^{-2\lambda \sigma^- } )\,.
\label{whole}
\end{eqnarray}
In terms of the space and the time coordinates the mirror trajectory is
given by
\begin{equation}
\sigma^{1} = \frac{1}{2\lambda}\,\ln \left[ \sqrt{1+(\frac{M}{4\lambda})^{2}
\,e^{4\lambda\sigma^{0}}\,} - \frac{M}{4\lambda}\,
e^{2\lambda\sigma^{0}} \: \right] \equiv z(\sigma^{0}) \,. \label{whtraject}
\end{equation}
Asymptotically the mirror approaches the null line,
\( \:
\sigma^{+} = x_{H}
\:, \) at an exponential rate of
\( \:
e^{-2\lambda \sigma^- }
\: \) as \( \:
\sigma^{-} \rightarrow \infty
\: \);
\begin{equation}
\sigma^{+} = p(\sigma^{-}) \: \rightarrow \: x_{H} - \frac{1}{M}\,
e^{-\,2\lambda\sigma^{-}} \,.
\end{equation}
Thus the parameters in the previous general discussion are given by
\begin{equation}
\kappa = 2\lambda \,, \hspace*{0.5cm} \xi = \frac{1}{M} \,,
\end{equation}
in this model.

Summarizing the property of the mirror trajectory necessary for the
discussion of event horizon formation, one demands the asymptotic form,
\( \:
p(\sigma^{-}) \rightarrow  x_{H} - \xi \,e^{-\,\kappa \sigma^{-}}
\: \) as
\( \:
\sigma^{-} \rightarrow \infty \,,
\: \)
and the initial form,
\( \:
p(\sigma^{-}) \rightarrow  \sigma^{-}
\: \) as $\sigma^{-} \rightarrow  -\,\infty$.
To asymptotic observers far away from the collapsing body the relevant
flat coordinate is the mirror coordinate $\sigma^{\pm }$ ;
\( \:
ds^{2} = -\,d\sigma^{+}d\sigma^{-}
\: \).

\addtocounter{chapter}{1}
\setcounter{equation}{0}
\section*{\lromn
3.
Quantum field with moving boundary and \\
Hawking radiation
}

\vspace{0.5cm}
\hspace*{0.5cm}
We now recapitulate the behavior of massless quantum field with the reflection
boundary. This subject is well developed, at least for the two spacetime
dimension, as reviewed for instance by the book of Birrell and Davies
\cite{bdavies}, but
we shall summarize the main points for the sake of completeness and also
because this is important to the following discussion.
Those who are familiar with this subject can skip this section and
can go to the next section immediately.
In this and the following sections we sometimes denote the mirror trajectory by
$z^{\pm }(\tau )$ with $\tau $ an arbitrary parameter specifying the path.
The entire half Minkowski spacetime is given, in terms of the
light cone coordinates $\sigma^{\pm }$, by
\( \:
\sigma^{+} \geq  p(\sigma^{-})
\: \).
The trajectory function previously defined by
$z ^{+} = p(z^{-})$ plays important roles in virtually all places in
the following discussion.

Massless quantum fields in two dimensions obey the simple
free field equation in the conformal gauge of
\( \:
ds^{2} = -\,e^{2\rho }\,d\sigma^{+}d\sigma^{-}
\: \);
\begin{equation}
\partial_{+}\partial_{-}\,f(\sigma^{+},\sigma^{-}) = 0 \,,
\end{equation}
with
\( \:
\partial_{\pm } = \frac{\partial }{\partial x^{\pm } } \,.
\: \)
This equation is valid even for  nontrivial cases of nonvanishing
curvature.
The positive frequency left-moving mode functions that obey
the reflection boundary condition are given by
\begin{equation}
f_{\omega}(\sigma^{+}\,, \sigma^{-}) = \frac{1}{\sqrt{2\omega}}\,[\:
e^{-i\omega\sigma^{+}} - e^{-i\omega p(\sigma^{-})}\:] \,,
\end{equation}
with $\omega > 0$ : \( \:
f_{\omega}(z^{+}(\tau ) \,, \,z^{-}(\tau )) = 0
\: \).
We may call these modes in-modes. Since in-modes and
their conjugates span a complete set of wave functions with the boundary
condition, one may expand the massless quantum field as
\begin{equation}
f(\sigma^{+},\sigma^{-})
= \frac{1}{2\pi}\, \int_{0}^{\infty}\,d\omega\,
(\,f_{\omega}a_{\omega}+f_{\omega}^{*}a_{\omega}^{\dag}\,) \,. \label{infield}
\end{equation}
The operators,
\( \:
a_{\omega} \:\mbox{and}\: a_{\omega}^{\dag} \,,
\: \)
are interpretated as annihilation and creation operators appropriate to
describe incoming plane waves onto the mirror.

Description of outgoing waves are less trivial. One might attempt to expand
in terms of a similar set of out-mode fuctions with the unit outgoing flux;
\begin{equation}
\tilde{f}_{\omega}(\sigma^{+} \,, \sigma^{-}) = \frac{1}{\sqrt{2\omega}}\,[\:
-\,e^{-i\omega\sigma^{-}} + e^{-i\omega q(\sigma^{+})}\:] \,.
\end{equation}
Here
\( \:
q(\sigma^{+})
\: \)
is defined by inverting the relation,
\( \:
z^{+} = p(z^{-})
\: \), with the result
\( \:
z^{-} = q(z^{+})
\: \)
such that the reflection condition,
\( \:
\tilde{f}_{\omega }(z^{+}(\tau )\,,\,z^{-}(\tau )) = 0
\: \),
is automatically satisfied.
Special care should however be taken in dealing with the spacetime endowed with
event horizon. The corresponding mirror trajectory $p(z^{-})$ is
bounded from above :
\( \:
p(z^{-}) \rightarrow x_{H}
\: \),
 and the inversion of this function is impossible for a large
enough $z^{+}$. Physically this means that incoming light rays
emitted very late never reach the future null infinity $I_{R}^{+}$ in the
right region as shown in Fig. 1.
The future null infinity thus splits into two parts, $I_{R}^{+}$ and
$I_{L}^{+}$, and out-modes  such as the previous $\tilde{f}_{\omega}$
relevant to observers at $I_{R}^{+}$ are not complete in
\( \:
I_{L}^{+} \cup I_{R}^{+}
\: \);
one must supplement these by another set of left moving out-modes,
$f_{\omega }^{L}$. In this general situation one may expand the field
in terms of previous \( \:
\tilde{f}_{\omega} \equiv f_{\omega }^{R}
\: \) and $f_{\omega }^{L}$ as
\begin{eqnarray}
f(\sigma^{+},\sigma^{-})
= \frac{1}{2\pi}\, \sum_{I=R,L}\,\int_{0}^{\infty}\,d\omega\,
(\,f_{\omega}^{I}\,a_{\omega}^{I}+f_{\omega}^{I *}\,a_{\omega}^{I\dag}\,) \,.
\label{outfield}
\end{eqnarray}

In the special example given previously by Eq.\ref{whole},
the right moving out-modes
\( \:
f_{\omega }^{R}
\: \)
are given, with
\begin{equation}
q(\sigma^{+}) =
-\,\frac{1}{2\lambda} \ln (e^{-2\lambda\sigma^+} - \frac{M}{2\lambda})
\, \,,
\end{equation}
by
\begin{eqnarray}
f^{R}_{\omega}
= \frac{1}{\sqrt{2\omega}}\,[\: -\,e^{-i\omega\sigma^{-}}
+ (e^{-2\lambda\sigma^{+}}-\frac{M}{2\lambda})^{i\frac{\omega}{2\lambda}}\:
\theta(x_{H} - \sigma^{+})\:] \,.
\end{eqnarray}

The two descriptions, either on the null past infinity $I^{-}$ by
Eq.\ref{infield}, or on the future
\( \:
I^{+}_{L}\cup I^{+}_{R}
\: \) by Eq.\ref{outfield},
are related by a Bogoliubov transformation of the form
\begin{eqnarray}
a_{\omega}
&=& \frac{1}{2\pi}\, \sum_{I=R,L}\,\int_{0}^{\infty}\,d\omega'\,
(\alpha_{\omega\omega'}^{I}\,a_{\omega'}^{I}+
\beta_{\omega\omega'}^{I}\,a_{\omega'}^{I \dag }) \,, \\
a_{\omega}^{I}
&=& \frac{1}{2\pi}\,\int_{0}^{\infty}\,d\omega'\,
(\tilde{\alpha}_{\omega'\omega}^{I}\,a_{\omega'}+
\tilde{\beta}_{\omega'\omega}^{I}\,a_{\omega'}^{\dag }) \,.
\end{eqnarray}
The Bogoliubov coefficients are defined by
\begin{eqnarray}
\alpha_{\omega\omega'}^{I} &=&
i\,\int_{-\infty}^{\infty}\,d\sigma^{+}\,f^{*}_{\omega}\,\tilde{\partial}_{+}
f_{\omega'}^{I}
\,, \hspace{0.5cm}
\beta_{\omega\omega'}^{I} =
i\,\int_{-\infty}^{\infty}\,d\sigma^{+}\,f^{*}_{\omega}\,\tilde{\partial}_{+}
f_{\omega'}^{I *} \,, \\
\tilde{\alpha}_{\omega'\omega}^{I} &=&
i\,\int_{I^{+}_{L}\cup I^{+}_{R}}\,d\sigma^{\mu } \,f^{I *}_{\omega}\,
\tilde{\partial}_{\mu }f_{\omega'}
\,, \hspace{0.5cm}
\tilde{\beta}_{\omega'\omega}^{I} =
i\,\int_{I^{+}_{L}\cup I^{+}_{R}}\,d\sigma^{\mu }\,f^{I *}_{\omega}\,
\tilde{\partial}_{\mu }f_{\omega'}^{*} \,,
\end{eqnarray}
with the Klein-Gordon inner product
\begin{equation}
i\,\int\,d\sigma^{\mu }\,
f_{1}^{*}\tilde{\partial}_{\mu } f_{2} \: \equiv \:
i\,\int\,d\sigma^{\mu }\,[\:f_{1}^{*}\partial_{\mu }f_{2} -
f_{2}\partial_{\mu }f_{1}^{*}\:]
\,. \label{kgproduct}
\end{equation}
Some exact symmetry and the completeness relations are written in Appendix
A and B,
together with explicit formulas of the Bogoliubov coefficients
for the previous specific model of Eq.\ref{whole}.

The crucial point to particle production is whether mode mixing between
positive and negative energy states occurs. This is judged whether
or not a type of Bogoliubov coefficients vanish;
\( \:
\tilde{\beta}_{\omega'\omega}^{I} = 0
\: \).
Indeed, at the right null infinity $I_{R}^{+}$, the particle number
produced from the in-vacuum is the expectation value of the particle
number operator computed in the in-vacuum;
\begin{equation}
\langle\, a_{\omega}^{R \dag}a_{\omega'}^{R}\, \rangle \,_{\rm{in}}
 =  \frac{1}{2\pi}\,\int_{0}^{\infty}\,d\omega''\,
\tilde{\beta}_{\omega''\omega}^{R *}\,\tilde{\beta}_{\omega''\omega'}^{R} \,.
\end{equation}
This formula may contain, as in the case of Hawking radiation, the factor
\( \:
2\pi \delta(\omega-\omega')
\: \), which expresses that the produced particle number is in proportion to
the duration of time, hence the total number integrated over an infinite time
interval is formally divergent.

In more intuitive physical terms this formula may be understood as
follows. Consider the in-mode function that obeys the reflection boundary
condition,
\begin{equation}
f_{\omega}(\sigma^{+},\sigma^{-}) = \frac{1}{\sqrt{2\omega}}\,
[\:e^{-i\omega\sigma^{+}} - e^{-i\omega p(\sigma^{-})}\:] \,.
\end{equation}
One may regard the two terms in the right hand side as the incident and the
reflected waves from the mirror, respectively.
If one computes according to the usual
flux formula the reflected flux, one finds that the flux is reduced by a
factor,
\begin{equation}
p'(z^{-}(\tau) ) = \frac{1+\frac{\partial z^{1}}{\partial z^{0}}}
{1-\frac{\partial z^{1}}{\partial z^{0}}} =
\frac{1+v}{1-v} \,,
\end{equation}
with $v$ the instantaneous velocity,
right at the mirror point. This is precisely the red shift factor caused by
the moving mirror. Namely, the flux is reduced, not necessarily because
the wave amplitude is diminished, but also because of the Doppler effect.
For the mirror moving with a constant velocity this is easy to understand;
with a constant $p'$ the reflected wave contains a single Doppler-shifted
component of frequency \( \:
\omega p'
\: \).
But with acceleration of the mirror something unusual happens;
the function in the exponent
\( \:
\omega p(\sigma^{-})
\: \) may
contain various Fourier modes when decomposed. In particular it may contain
negative frequency components. In fact, the mixing amplitude with a negative
frequency is given by
\( \:
\tilde{\beta}_{\omega \omega'}^{R*}
\: \): the Bogoliubov coefficient. Thus, according to the hole theory idea,
the number of emitted particles upon reflection is the number of mixed
negative energy states that have been occupied before reflection,
\begin{equation}
\frac{1}{2\pi}\,\int_{0}^{\infty}\,d\omega''\,
\tilde{\beta}_{\omega''\omega}^{R *}\,\tilde{\beta}_{\omega''\omega}^{R}
 = \langle\, a_{\omega}^{R \dag}a_{\omega}^{R}\, \rangle \,_{\rm{in}} \,.
\end{equation}
This coincides with the previous calculation of produced particles.

The thermal spectrum of Hawking radiation follows when one takes the mirror
trajectory approaching a null line at the exponential rate. Using the
previous asymptotic form of the trajectory,
\( \:
p(\sigma^{-}) = x_{H} - \xi \,e^{-\,\kappa \sigma^{-}} \,,
\: \)
the Bogoliubov coefficient is computed as
\begin{eqnarray}
\tilde{\beta}_{\omega'\omega}^{R} &=&
i\,\int_{-\infty }^{\infty }\,d\sigma \frac{1}{\sqrt{4\omega \omega'}}\,
e^{i\omega \sigma }\,\tilde{\partial }\;\mbox{exp}[\,i\omega'(x_{H} -
\xi \,e^{-\,\kappa \sigma })\,]
\nonumber \\ &=&
\frac{1}{\kappa }\,\sqrt{\frac{\omega}{\omega'}}\,
e^{i\omega' x_{H}}\,
(\xi \omega')^{i\frac{\omega}{\kappa }}\,
e^{-\frac{\pi\omega}{2\kappa }}\:
\Gamma(-i\,\frac{\omega}{\kappa })  \,. \label{bogoliubovb}
\end{eqnarray}
This yields the particle number observed on $I_{R}^{+}$ at late times;
\begin{eqnarray}
\langle\, a_{\omega}^{R \dag}a_{\omega'}^{R}\, \rangle \,_{\rm{in}}
&=& \frac{1}{\kappa^{2}}\,\sqrt{\omega \omega'}\,e^{-\,\frac{\pi }{2\kappa }
(\omega + \omega')}\:\Gamma(i\,\frac{\omega }{\kappa })\,
\Gamma(-i\,\frac{\omega' }{\kappa })\,\frac{1}{2\pi }\,\int_{0}^{\infty }\,
\frac{d\omega''}{\omega''}\,(\xi \omega'')^{\frac{i}{\kappa }(\omega' -
\omega )}
\nonumber \\
&=&  2\pi\,\delta(\omega-\omega')\:
\frac{1}{e^{\frac{2\pi\omega}{\kappa}}-1} \,.
\end{eqnarray}
It is precisely of the form of Planck distribution with the temperature
\( \:
T = \frac{\kappa }{2\pi}
\: \).

One may also compute correlation functions and the stress tensor, which
becomes important in the later discussion of back reaction.
It is straightforward to calculate the two point correlation function
by using the in-mode expansion. The result is
\begin{eqnarray}
\langle\, f(\sigma)f(\sigma')\, \rangle \,_{\rm{in}} &=&
\frac{1}{2\pi}\,\int_{0}^{\infty}\,d\omega\,
\frac{1}{2\omega }\,[\:e^{-\,i\omega \sigma^{+}} -
e^{-\,i\omega p(\sigma^{-})}\:]\,
[\:e^{i\omega \sigma'^{+}} - e^{i\omega p(\sigma'^{-})}\:]
\nonumber \\ &=&
-\,\frac{1}{4\pi}\,\ln \frac{[\:\sigma^{+}-\sigma'^{+}-i\epsilon\:]\,
[\:p(\sigma^{-})-p(\sigma'^{-})-i\epsilon\:]}
{[\:\sigma^{+}-p(\sigma'^{-})-i\epsilon\:]\,
[\:p(\sigma^{-})-\sigma'^{+}-i\epsilon\:]} \,,
\end{eqnarray}
with \( \:
\epsilon \rightarrow 0^{+}
\: \) in the end of calculation.
In the case of the fixed boundary this reduces to
\begin{equation}
-\,\frac{1}{4\pi}\,\ln \frac{[\:\sigma^{+}-\sigma'^{+}-i\epsilon\:]\,
[\:\sigma^{-} - \sigma'^{-} - i\epsilon\:]}
{[\:\sigma^{+} - \sigma'^{-}-i\epsilon\:]\,
[\:\sigma^{-} - \sigma'^{+}-i\epsilon\:]} \,,
\end{equation}
with \( \:
p(\sigma^{-}) = \sigma^{-}
\: \).
Observable correlations of the massless field are given by
\begin{eqnarray}
&&
\langle\, \partial_{-}f(\sigma)\partial_{-}f(\sigma')\, \rangle\,_{\rm{in}} =
-\,\frac{1}{4\pi}\,\frac{p'(\sigma^{-})\,p'(\sigma'^{-})}
{[\,p(\sigma^{-}) - p(\sigma'^{-}) - i\epsilon\,]^{2}} \,, \\
&&
\langle\, \partial_{+}f(\sigma)\partial_{+}f(\sigma')\, \rangle \,_{\rm{in}}
 =
-\,\frac{1}{4\pi}\,\frac{1}{(\sigma^{+} - \sigma'^{+} - i\epsilon)^{2}}
 \,, \\
&&
\langle\, \partial_{-}f(\sigma)\partial_{+}f(\sigma')\, \rangle \,_{\rm{in}}
 =
-\,\frac{1}{4\pi}\,\frac{p'(\sigma^{-})}
{[\,p(\sigma^{-}) - \sigma'^{+} - i\epsilon\,]^{2}}
 \,.
\end{eqnarray}

Conceptually it is better to distinguish correlators among right movers from
those of the entire field $f$. By disentangling the left movers, we may
define the quantum field $f^{R}$ relevant to asymptotic observers at
$I_{R}^{+}$ consisting of one of the two sets of modes
\( \:
f_{\omega}^{R}
\: \);
\begin{eqnarray}
f^{R}(\sigma^{+},\sigma^{-})
= \frac{1}{2\pi}\, \int_{0}^{\infty}\,d\omega\,
(f_{\omega}^{R}\,a_{\omega}^{R}+f_{\omega}^{R *}\,a_{\omega}^{R\dag}) \,.
\end{eqnarray}
In principle it should be possible to construct the density matrix relevant
to observers at $I_{R}^{+}$ from the field defined this way.
Ignoring irrelevant infinities and writing the correlator valid
in the region away from the horizon,
\( \:
\sigma^{+} > x_{H}
\: \),
if it exists,
\begin{eqnarray}
\langle\, f^{R}(\sigma)f^{R}(\sigma')\, \rangle \,_{\rm{in}} =
-\,\frac{1}{8\pi}\, \ln\frac{
[\,p(\sigma^{-}) - p(\sigma'^{-}) - i\epsilon\,]^{2}}
{[\,x_{H} - p(\sigma'^{-}) - i\epsilon\,]\,
[\,p(\sigma^{-}) - x_{H} - i\epsilon\,]} \,.
\end{eqnarray}
Derivation of this and related formulas is sketched in Appendix C.
Some of our discussion is simplified by using the identity among the two
field operators,
\( \:
\partial_{-}f = \partial_{-}f^{R}
\: \). This operator identity should be evident since the right movers consist
of those field
components having reflected waves. The two descriptions, either in terms of
$f$ or $f^{R}$, must give the identical result if one restricts observations
to those at $I_{R}^{+}$.
Thus it is possible to explicitly verify that
\begin{equation}
\langle\, \partial_{-}f^{R}(\sigma)\partial_{-}f^{R}(\sigma')
\, \rangle \,_{\rm{in}} =
\langle\, \partial_{-}f(\sigma)\partial_{-}f(\sigma')\, \rangle \,_{\rm{in}}
 \,.
\end{equation}

It is useful to write the two point correlator of the right moving field
in terms of the vacuum expectation values in the in-vacuum.
Using the out-mode completeness,
\begin{equation}
\frac{1}{2\pi}\, \int_{0}^{\infty}\,d\omega^{''}\,
[\,\tilde{\alpha}_{\omega''\omega'}^{R}\,\tilde{\alpha}_{\omega''\omega}^{R *}-
\tilde{\beta}_{\omega''\omega'}^{R}\,\tilde{\beta}_{\omega''\omega}^{R *}\,]
= 2\pi\,\delta(\omega-\omega')\,,
\end{equation}
one may eliminate
\( \:
\tilde{\alpha }\,\tilde{\alpha }^{*}
\: \) terms, hence \( \:
\langle\, a^{R}a^{\dag R}\, \rangle\:_{{\rm in}}
\: \), in the two point correlator. Thus one derives
\begin{eqnarray}
&&
\langle\, f^{R}(\sigma)f^{R}(\sigma')\, \rangle\,_{\rm{in}} =
\int_{0}^{\infty}\, \frac{d\omega}{2\pi}\,
f_{\omega}^{R}(\sigma )\,f_{\omega}^{R*}(\sigma') +
\int_{0}^{\infty}\, \frac{d\omega}{2\pi}\,
\int_{0}^{\infty}\, \frac{d\omega'}{2\pi}\,
\nonumber \\
&&
\left(
2\,\mbox{Re}[\,f_{\omega}^{R*}(\sigma )f_{\omega'}^{R}(\sigma')\, \langle\,
a_{\omega }^{\dag R}a_{\omega'}^{R}\, \rangle\,_{\rm{in}}\,] +
2\,\mbox{Re}[\,f_{\omega}^{R}(\sigma )f_{\omega'}^{R}(\sigma')\, \langle\,
a_{\omega }^{R}a_{\omega'}^{R}\,\rangle\,_{\rm{in}}\,]
 \right) \,. \nonumber \\ \label{correlationg}
\end{eqnarray}
In this formula the first term represents the free field contribution with
the fixed boundary, and one may look into absolute and relative significance
of deviation from this, the last two terms,
in a variety of situations of given mirror trajectories.

We now discuss the quantum stress tensor
\( \:
\langle\, T_{\pm \pm }\, \rangle
\: \); the expectation value computed in the in-vacuum.
Its evaluation involves the coincidence limit that diverges.
Precisely which operator is relevant in the back reaction right at the mirror
is discussed in later sections. Here is derived the asymptotic form of the
stress tensor observable at the past and the future infinities.
For this purpose we take the point splitting regularization,
\begin{equation}
\lim_{\sigma' \rightarrow \sigma}
\partial_{\pm}\,\partial_{\pm}'\,\langle\, f(\sigma)f(\sigma')\, \rangle\,
_{\rm{in}} \,,
\end{equation}
and subtract the coincidence limit of field correlaters in the case of
the fixed boundary;
thus the local quantum stress tensor is defined by
\begin{equation}
\langle \,T_{\pm \pm }(\sigma ) \, \rangle = \lim_{\sigma' \rightarrow \sigma}
\partial_{\pm }\,\partial_{\pm }'\,\langle\, f(\sigma)f(\sigma')\, \rangle\,
_{\rm{in}} -
(-\,\frac{1}{4\pi})\,\lim_{\sigma' \rightarrow \sigma}
\frac{1}{(\,\sigma^{\pm}-\sigma'^{\pm}-i\epsilon\,)^{2}} \,.
\end{equation}
The only nonvanishing stress tensor component is then
\begin{equation}
\langle\, T_{--}(\sigma)\, \rangle =
\frac{1}{12\pi}\,(\partial_{-}p)^{1/2}\,\partial_{-}^{2}(\partial_{-}p)^{-1/2}
= -\,\frac{1}{24\pi}\,\{ p\,,\: \sigma^{-} \}_{S} \,. \label{aflux}
\end{equation}
Here \( \:
\{\:\,,\: \}_{S}
\: \)
is the Schwartzian derivative whose definition and some of their simple
properties are given in Appendix D.
It turns out, as will be discussed in the next section, that in the simplest
model the tensor component Eq.\ref{aflux} reduced by a factor $2$
is the one appearing in the back reaction formula.
It should be evident that the stress tensor computed this way
correctly describes the asymptotic flux at the null infinity \( \:
I_{R}^{+}
\: \) if one starts from the vacuum state at the past null infinity $I^{-}$.

Another form of the asymptotic flux invites an illuminating interpretation.
In terms of the velocity \( \:
v = \frac{\partial z^{1}}{\partial z^{0}}
\: \) of the mirror particle,
\( \:
\partial_{-}p = \frac{1+v}{1-v}
\: \), and
\begin{equation}
\langle\, T_{--}(\sigma)\, \rangle =
-\,\frac{1}{12\pi}\,\frac{(1+v)^{2}}{(1-v^{2})^{3/2}}\:\frac{d}{dt}
\frac{v'}{(1-v^{2})^{3/2}} \,,
\end{equation}
with \( \:
t = z^{0}
\: \) and the prime denoting the $t-$derivative.
The acceleration $\alpha $ in the instantaneous rest frame
may be defined as
\begin{equation}
\alpha = \frac{d}{dt}\,\frac{v}{\sqrt{1-v^{2}}} = \frac{v'}
{(1-v^{2})^{3/2}} \,. \label{acceleration}
\end{equation}
Thus \( \;
\langle\, T_{--}\, \rangle \propto -\,\frac{d}{dt}\,\alpha
\: \). It might sound strange, but is true that a uniformly accelerated mirror
does not produce energy flux in the asymptotic region.

It is now appropriate to discuss the nature of correlation when a particular
form of the mirror trajectory $p(\sigma^{-})$ is given.
First, for a constant mirror velocity,
\( \:
p(\sigma^{-}) = p_{1}\,\sigma^{-}
\: \) with a constant $p_{1} \neq 0$, and the correlation function
\( \:
\langle\, \partial_{\pm }f(\sigma)\partial_{\pm }f(\sigma')\, \rangle \,
_{\rm{in}}
\: \) coincides with the free field correlation in the fixed boundary case.
The stress tensor also coincides with that of the fixed boundary case;
\( \:
\langle\, T_{--}(\sigma)\, \rangle = 0
\: \).
Thus there is no sign of the breakdown
of quantum mechanical evolution if the end point of Hawking evaporation is
the moving mirror receding with a constant velocity. This should be so,
because there is no
mixing of positive and negative energy states in this case, as discussed
previously.

On the other hand, if one takes the accelerated trajectory
towards the null line with the exponential rate
\( \:
\sigma^{+} = x_{H} - \xi \,e^{-\,\kappa \sigma^{-}}
\: \)
, one finds that
\begin{eqnarray}
&&
\langle\, \partial_{-}f(\sigma)\partial_{-}f(\sigma')\,\rangle \,_{\rm{in}}
=
-\,\frac{\kappa^{2}}{16\pi}\,\frac{1}{\sinh^{2}\Delta} \,, \hspace{0.5cm}
\Delta = \frac{\kappa }{2}\,(\sigma^{-} - \sigma'^{-})
\,, \\
&&
\langle\, T_{--}(\sigma)\, \rangle =
\frac{\kappa^{2}}{48\pi}\,. \label{aaflux}
\end{eqnarray}
The right moving out-mode decomposition is diagonal in the sense that
\( \:
\langle\, a_{\omega }^{R \dag}a_{\omega'}^{R}\, \rangle \,_{\rm{in}}
\propto
\delta (\omega - \omega')
\: \)
and
\begin{equation}
\langle\, a_{\omega }^{R}a_{\omega'}^{R}\, \rangle\,_{\rm{in}} =
\frac{1}{2\pi }\,\int_{0}^{\infty }\,d\omega''\:
\tilde{\alpha}_{\omega''\omega }^{R}\,\tilde{\beta}_{\omega''\omega'}^{R} \:
\propto \:
\frac{1}{\sqrt{\omega \omega'}}\,\delta (\omega + \omega') \,,
\end{equation}
which may effectively be regarded as zero when one considers observable
quantities by multiplying $\omega \omega'$. An expression for
\begin{equation}
\tilde{\alpha}_{\omega'\omega }^{R} =
\frac{1}{\kappa }\,\sqrt{\frac{\omega}{\omega'}}\,
e^{-\,i\omega' x_{H}}\,
(\xi \omega')^{i\frac{\omega}{\kappa }}\:
e^{\frac{\pi\omega}{2\kappa }}\:
\Gamma(-i\,\frac{\omega}{\kappa })  \,, \label{bogoliubova}
\end{equation}
together with
\( \:
\tilde{\beta }_{\omega' \omega }^{R}
\: \) in Eq.\ref{bogoliubovb},
was used in this derivation.
All this indicates the thermal behavior with a temperature
\( \:
T = \frac{\kappa }{2\pi}
\: \).
The constant energy flux arises from that the deceleration of the mirror
particle is very large,
\begin{equation}
\alpha =
\frac{v'}{(1-v^{2})^{3/2}} = -\,\frac{1}{2}\,\sqrt{\frac{\kappa }{\xi}}
\,e^{\frac{1}{2}\,\kappa \sigma^{-}} \,,
\end{equation}
in sharp contrast to the case of the uniform acceleration.

We however note that the thermal behavior is not the whole story.
As stresssed in Ref.\cite{carlitz}, \cite{wilcz93},
there is a correlation among the left and the right movers as seen in
\begin{equation}
\langle\, \partial_{-}f(\sigma)\partial_{+}f(\sigma')\,
\rangle \,_{\rm{in}} \sim
-\,\frac{1}{4\pi }\,\frac{\xi \kappa e^{-\,\kappa \sigma^{-}}}
{(x_{H} - \xi \,e^{-\,\kappa \sigma^{-}}- \sigma'^{+})^{2}} \,,
\end{equation}
which becomes appreciable as $\sigma'^{+}$ approaches the horizon
at $x_{H}$. The correlation may get quite strong as
\( \:
\sigma'^{+} \rightarrow  x_{H}
\: \) and in the limit
\begin{equation}
\langle\, \partial_{-}f(\sigma)\partial_{+}f(\sigma')\, \rangle \,_{\rm{in}}
\rightarrow
-\,\frac{1}{4\pi }\,\frac{\kappa }{\xi}\,e^{\kappa \sigma^{-}} \,.
\label{thlimit}
\end{equation}
In terms of modes one can observe a sharper correlation \cite{carlitz} .
For this purpose define a convenient set of left moving out-modes by
\begin{equation}
f_{\omega }^{L} = \frac{1}{\sqrt{2\omega }}\,
[\:\frac{\sigma - x_{H}}{\xi }\:]^{-\,i\frac{\omega}{\kappa }}\,
\theta (\sigma - x_{H}) \,.
\end{equation}
Using
\begin{eqnarray}
\tilde{\beta}_{\omega'\omega}^{L} &\sim&
\frac{i}{\sqrt{4\omega \omega'}}\,\int_{x_{H}}^{\infty }\,d\sigma \,
[\:\frac{\sigma - x_{H}}{\xi }\:]^{i\frac{\omega}{\kappa }}\:
\tilde{\partial }\,e^{i\omega'\sigma } \nonumber \\
&=&
\frac{1}{\kappa }\,\sqrt{\frac{\omega}{\omega'}}\,e^{i\omega'x_{H}}\,
(\xi \omega')^{-\,i\frac{\omega}{\kappa }}\:
e^{-\frac{\pi\omega}{2\kappa }}\:
\Gamma(i\,\frac{\omega}{\kappa })  \,,
\end{eqnarray}
together with the expression for
\( \:
\tilde{\alpha }_{\omega'\omega}^{R}
\: \) Eq.\ref{bogoliubova},
one obtains a correlation between the left and the right modes,
\begin{eqnarray}
\langle\, a_{\omega}^{R }a_{\omega'}^{L}\, \rangle\,_{\rm{in}}
& = & \frac{1}{2\pi}\,\int_{0}^{\infty}\,d\omega''\:
\tilde{\alpha }_{\omega''\omega}^{R}\,\tilde{\beta}_{\omega''\omega'}^{L}
\nonumber \\
& \sim & \delta(\omega-\omega')\,\frac{\pi}{\sinh\frac{\pi\omega}{\kappa }}
 \,.
\end{eqnarray}

Moreover, a possible subleading term to the asymptotic formula
\( \:
p(\sigma^{-}) = x_{H} - \xi \,e^{-\,\kappa \sigma^{-} }
\: \)
may indicate a subtle non-thermal correlation.
In other words, the thermal behavior is a direct consequence of the
exponential approach to a null line in the mirror picture, or equivalently
the existence of the global event horizon in the collapse spacetime.
With the effect of back reaction included it is not clear that this
behavior of the mirror is maintained. Importance of the back reaction
is here, in elucidating the end point behavior of the mirror trajectory.

\addtocounter{chapter}{1}
\setcounter{equation}{0}
\section*{\lromn
4. Dynamical moving mirror with quantum back reaction
}

\vspace{0.5cm}
\hspace*{0.5cm}
So far we discussed the situation in which the mirror trajectory is
specified by hand, or the background metric is fixed in the collapse
spacetime. As the next and the most crucial step we wish to
incorporate the back reaction caused by emission of particles.
In this circumstance one has to elevate the mirror variable $z^{\pm }(\tau )$
to a dynamical one. This is most straightforwardly achieved by going to
the action principle. This procedure should in principle be able to
deduce the correct form of the back reaction, but the
boundary term in the field action must be treated with great care.

Suppose that the classical part of the mirror action is given by
\begin{equation}
S_{m} = \int_{-\,\infty }^{\infty }\,d\tau \, l(z^{\pm },\dot{z}^{\pm })\,,
\end{equation}
with the dot meaning \( \:
\frac{d}{d\tau }
\: \) hereafter.
As in the case of the usual point mass we demand that this lagrangian is
invariant under reparametrization of the arbitrary parameter $\tau $
specifying the mirror path,
\begin{equation}
\tau \: \rightarrow \: \tau' = \tau'(\tau) \,.
\end{equation}
Requirement of the reparametrization
invariance implies a constraint on the boundary term that arises from the
field variation; when one writes this variation as
\begin{equation}
\delta_{b} S = \int\,d\tau \, (F_{+}\delta z^{+} + F_{-}\delta z^{-}) \,,
\end{equation}
the condition
\begin{eqnarray}
F_{+}\,\dot{z}^{+} + F_{-}\,\dot{z}^{-} = 0 \,, \label{fconstraint}
\end{eqnarray}
must follow. Actually, this equality holds for each piece of
the action respecting the reparametrization invariance.
Otherwise there should not be much constraint on the form of the mirror
action, except that the given action should describe the behavior
of the underlying dynamics of gravity. How to choose this classical action
will be discussed in the next section. In this section we shall deal with
the back reaction term, first clarifying it classically.

Consider the action of a massless field with the boundary at the mirror,
\begin{eqnarray}
S_{f}
&=&
\int_{\sigma^{+} > z^{+}(\sigma^{-})}\,
d\sigma^{+}d\sigma^{-}\,\partial_{+}\,f\partial_{-}\,f
 \nonumber \\
&=&
\int_{-\,\infty }^{\infty }\,
d\sigma^{+}d\sigma^{-}\,\partial_{+}\,f\partial_{-}\,f
\:\theta (\sigma^{+} - z^{+}(\sigma^{-})) \nonumber \\
&=&
\int_{-\,\infty }^{\infty }\,
d\sigma^{+}d\sigma^{-}\,d\tau \:\partial_{+}\,f\partial_{-}
\,f\:\theta (\sigma^{+} - z^{+}(\sigma^{-})\,)\,\dot{z}^{-}\delta(\sigma^{-} -
z^{-}(\tau )\,) \,.
\end{eqnarray}
We introduced the same arbitrary parameter $\tau $ as in the mirror action.
Considering both the field and the boundary variation, one is led to
\begin{eqnarray}
\delta S_{f} &=& -\,2\,
\int_{\sigma^{+} > z^{+}(\sigma^{-})}\,
d\sigma^{+}\,d\sigma^{-}\: \partial_{+}\partial_{-}\,f\,\delta f
\nonumber  \\ &&
+ \int_{-\,\infty }^{\infty }\,d\tau \,
[\: (\dot{z}^{+}\delta z^{-}
- \dot{z}^{-}\delta z^{+})\, \partial_{+}\,f\partial_{-}\,f
- (\dot{z}^{-}\partial_{-}\,f - \dot{z}^{+}\partial_{+}\,f)\,\delta f \:] \,.
\end{eqnarray}
The first term gives the usual field equation, while the second contains
the boundary variation $\delta z^{\pm }$
as well as the surface term upon partial integration.
No use of the boundary condition
\begin{equation}
f(z^{+}(\tau)\,, \,z^{-}(\tau)) = 0 \,,
\end{equation}
is made in this derivation.
Thus both for the Dirichlet and the Neumann condition, the boundary
term arising from the variational principle is
\begin{equation}
\delta_{b} S_{f} =
-\,\int\, d\tau\,
\left( T_{++}\,\dot{z}^{+}\delta z^{+} - T_{--}\,\dot{z}^{-}
\delta z^{-}\right)  \,, \label{bterm}
\end{equation}
with \( \:
T_{\pm \pm } = (\partial_{\pm }f)^{2}
\: \) the stress tensor components. This form automatically satisfies the
force constraint Eq.\ref{fconstraint}, since
\( \:
T_{++}(\dot{z}^{+})^{2} = T_{--}(\dot{z}^{-})
^{2}
\: \) at the classical level.

In order to clarify the nature of the back reaction against the mirror point,
let us assume for the moment the usual form of the action for the point mass,
\begin{equation}
S_{m} = -\,m\, \int\, d\tau\,\sqrt{\dot{z}^{+}\dot{z}^{-}} \,.
\end{equation}
Combined with the above form of the boundary interaction with the $f-$field,
the action principle leads to the equation of the mirror motion,
\begin{equation}
\frac{m}{2}\,\frac{d}{d\tau }\,\sqrt{\frac{\dot{z}^{\mp}}{\dot{z}^{\pm}}}
\pm T_{\pm \pm }\,\dot{z}^{\pm } = 0 \,,
\end{equation}
with $T_{\pm \pm }$ evaluated at the boundary. In the light cone gauge
of \( \:
\tau = z^{-}
\: \),
\begin{equation}
\frac{m}{4}\,\frac{\ddot{z}^{+}}{\sqrt{\dot{z}^{+}}} + T_{--} = 0 \,.
\end{equation}
This is the usual form of the momentum conservation; the total momentum
sum of the mirror point and the field back reaction is balanced.
The hamiltonian
identically vanishes by the reparametrization invariance. It is then
clear that the mirror point must have a negative mass,
\( \:
m < 0
\: \) in the notation here, in order to describe physically
the effect of the back reaction. With the conventional sign of a positive
mass the mirror point would be further accelerated to the left,
instead of the right, by emitting
particles; an unacceptable situation for description of the back reaction.
Hence we shall hereafter change $m$ to $-\,m$
such that the parameter $m$ is always positive.

The classical boundary term Eq.\ref{bterm} that expresses the back reaction
by incident and outgoing waves must be modified to incorporate the effect
of spontaneous particle emission such as the Hawking radiation. This
involves something like replacing the classical stress tensor by
\( \:
\langle \, T_{\mu \nu } \, \rangle \,,
\: \)
the expectation value of the stress tensor operator computed in the in-vacuum.
But fixing the local back reaction term precisely needs some careful
argument.
First, a well-defined result for the quantum stress tensor follows
in the asymptotic region far away from the mirror,
by subtracting the free field limit in the case of the fixed boundary
as done in the preceding section;
\begin{eqnarray}
\langle\,  T_{--}(\sigma^{-})\, \rangle  =
-\, \frac{1}{24\pi }\,\{ \sigma^{+}\,, \: \sigma^{-} \}_{S}
=
-\,\frac{1}{24\pi}\,
[\: \frac{\partial_{-}^{3}\sigma^{+}}{\partial_{-}\sigma^{+}}
- \frac{3}{2}\,(\frac{\partial_{-}^{2}\sigma^{+}}
{\partial_{-}\sigma^{+}})^{2} \:]\,. \label{asymflux}
\end{eqnarray}
Note that this form is manifestly reparametrization invariant at the
location of the mirror:
\( \:
\sigma^{\pm } = z^{\pm }(\tau ) \,.
\: \)
All the other $\langle\, T_{\mu \nu } \,\rangle $ \( \:
(\mu \,, \nu  = \pm )
\: \)
vanishes. This stress tensor represents the
asymptotic flux observed at the future null infinity $I_{R}^{+}$, when
the state at the past null infinity $I^{-}$ was the vacuum.

Reparametrization invariant quantum back reaction at the mirror
must obey the force constraint Eq.\ref{fconstraint}.
Assuming a symmetry
under \( \:
z^{+} \leftrightarrow z^{-}
\: \), we generalize the classical force \( \:
\mp \,T_{\pm \pm }\,\dot{z}^{\pm }
\:, \)
to a modified quantum force
\( \:
F_{\pm } = \mp\,\left( \langle \,T_{\pm \pm } \,\rangle +
G(z^{\mp }\,,z^{\pm}) \right)\,\dot{z}^{\pm }
\: \)
such that the reparametrization invariance takes the form of
\begin{equation}
\left(\,\langle\, T_{--}\, \rangle + G(z^{+},z^{-})\,
\right)\,(\dot{z}^{-})^{2} =
\left(\,\langle\, T_{++}\, \rangle + G(z^{-},z^{+})\,
\right)\,(\dot{z}^{+})^{2} \,.
\end{equation}
To some extent the choice of \( \:
G(z^{+},z^{-})
\: \) is arbitrary.
The consistency with the asymptotic flux
Eq.\ref{asymflux} for
\( \:
\langle\, T_{\mu \nu }\, \rangle \,,
\: \)
combined with a minimum number of the derivatives, suggests that
\begin{equation}
G(z^{+}\,,\,z^{-}) = \alpha\, [\: \{ z^{+}\,,\,z^{-}\}_{S} + \gamma\,
(\frac{\partial_{-}^{2}z^{+}}{\partial_{-}z^{+}})^{2}\:] \,,
\end{equation}
with $\alpha $ and $\gamma $ constants to be determined.
As a minimal requirement we demand that
the quantum \( \:
\langle\, T_{--}\, \rangle
\: \) should agree with the asymptotic flux Eq.\ref{asymflux}
when no incoming flux is present, \( \:
\langle\, T_{++}\, \rangle = 0 \,.
\: \)
From this consideration
one finds that $\alpha = \frac{1}{48\pi }$ unambiguously.
Thus the back reaction is given by
\begin{equation}
F_{\mp} =
-\,\frac{1}{48\pi }\,[\: \{ z^{\pm }\,,\: z^{\mp}\}_{S} \mp \gamma\,
(\frac{\partial_{\mp}^{2}z^{\pm }}{\partial_{\mp}z^{\pm }})^{2}\:]\,
\dot{z}^{\mp} \,,
\end{equation}
with a free, undetermined parameter $\gamma $.

Furthermore, as a physical requirement we demand that
a definite sign for the back reaction
should automatically follow when the event horizon is formed.
Namely, assuming the exponential approach to a null line for the mirror
trajectory,
\( \:
z^{+} \sim x_{H} - \xi\,e^{-\,\kappa z^{-}} \,,
\: \)
one derives that
\begin{equation}
F_{-} \sim \frac{1}{48\pi }\,(\frac{1}{2}+\gamma )\,\kappa^{2}\,
\frac{dz^{-}}{d\tau } \,,
\end{equation}
as \( \:
z^{-} \rightarrow \infty
\: \).
The correct sign for the particle emission then gives
\begin{equation}
\gamma \geq -\,\frac{1}{2} \,.
\end{equation}
Various choices of $\gamma $ are possible at this point ;
we mainly consider the case of
$\gamma = 0$.
Chung and Verlinde \cite{verlin93} took a different value of
$\gamma $.

There is a unique feature for the choice of $\gamma = 0$.
Suppose that we seek, as possible candidates of the end point
behavior, all solutions in which the quantum back reaction
does not contribute in the mirror equation, by imposing that
\( \:
F_{-} = 0
\: \). This immediately gives the asymptotic outgoing flux,
\begin{equation}
\langle\, T_{--}\, \rangle = -\,\frac{1}{24\pi }\,\{ z^{+}\,,\:z^{-}\}_{S} =
-\,\frac{\gamma}{24\pi}\,(\frac{\partial_{-}^{2}z^{+}}{\partial_{-}z^{+}})^{2}
\,.
\end{equation}
The positivity of the flux, \( \:
\langle\, T_{--}\, \rangle\, \geq  0
\: \), implies that $\gamma \leq  0$.
Let us now write the asymptotic flux in terms of the velocity $v$ of the mirror
particle. Noting that
\begin{equation}
\partial_{-}z^{+} = \frac{1+v}{1-v} \,,
\end{equation}
one finds with vanishing back reaction that
\begin{equation}
\langle\, T_{--} \rangle = -\,\frac{\gamma }{6\pi }\,
\frac{v'^{2}}{(1-v)^{2}\,(1-v^{2})^{2}} \,.
\end{equation}
Here the prime indicates the $z^{0}-$derivative.
This means that in order to get identically vanishing flux the mirror should
stop accelerating for $\gamma \neq 0$ :
\( \:
v' = 0
\: \).
For the special case of $\gamma = 0$
the condition of the vanishing back reaction against the mirror
implies automatically vanishing asymptotic flux at infinity.
Because of this nice feature we attach a special significance to this case.
Although we discuss the cases of $\gamma \neq 0$ in later sections, we first
concentrate on the special case of $\gamma =0$. Indeed, the end point behavior
is most sensible for this case, as will be born out by our analysis.

Summarizing this section, we write the dynamical equation for the mirror
particle without the reparametrization gauge fixing. With the classical
part of the mirror lagrangian given by
\( \:
m\,l(z^{\pm },\dot{z}^{\pm })
\: \), it has the form of
\begin{equation}
-\,m\,\frac{\delta }{\delta z^{\pm }}\, l(z^{\pm },\dot{z}^{\pm }) =
-\,\frac{1}{48\pi }\,[\: \{ z^{\mp }\,,\:z^{\pm}\}_{S} \pm
\gamma\,(\frac{\partial_{\pm}^{2}z^{\mp }}{\partial_{\pm}z^{\mp }})^{2}\:]\,
\dot{z}^{\pm} \,.
\end{equation}
Explicitly in terms of the parameter $\tau $,
\begin{equation}
\{ z^{+ }\,,\:z^{-}\}_{S} -
\gamma\,(\frac{\partial_{-}^{2}z^{+ }}{\partial_{-}z^{+ }})^{2} =
(\dot{z}^{-})^{-2}\:[\:
\frac{\stackrel{...}{z}^{\,+}}{\dot{z}^{+}} - \frac{3}{2}\,
(\frac{\ddot{z}^{+}}{\dot{z}^{+}})^{2}
- \frac{\stackrel{...}{z}^{\,-}}{\dot{z}^{-}} + \frac{3}{2}\,
(\frac{\ddot{z}^{-}}{\dot{z}^{-}})^{2} - \gamma \,
(\frac{\ddot{z}^{+}}{\dot{z}^{+}} - \frac{\ddot{z}^{-}}{\dot{z}^{-}})^{2}
\:] \,. \nonumber \\
\end{equation}
A similar expression follows with the interchange:
\( \:
z^{+} \: \leftrightarrow \: z^{-} \,.
\: \)

Prior to solving the back reaction problem in the next section,
it would be useful to classify possible types of end point behavior
of the  trajectory and its implications on Hawking evaporation
in the moving mirror picture.
Two extreme cases of the end point behavior are immediately listed:
complete stop and
continued approach to a null line at the exponential rate.
In the first case the end point spacetime is a flat
Minkowski space, and the information loss puzzle is resolved most
straightforwardly. Of course even in this case how the almost lost
information is recovered in the late phase of Hawking evaporation
is an interesting problem.
In the second case quantum mechanics is violated, but we may learn
how the drastic revision of the fundamental physical law
might be avoided by introducing new physics neglected in this approach.
Another less obvious, but interesting possibility is a final mirror motion
with a constant recession velocity. In this case quantum mechanics
is maintained, but the final state
contains an object that renders all incoming waves red-shifted by
a universal amount
\( \:
\frac{1+v}{1-v}
\: \)
determined by the final mirror velocity $v$.
One may justifiably call this object a remnant \cite{wilcz93}.
The remnant must have unusual properties as a source of gravity.
The possible existence
of remnants at the end point of Hawking evaporation has been discussed
many times in the literature \cite{remnant}.
There may be other possibilities that are not easily guessed at
this point. In any case the result will come out automatically
once the classical
mirror model is set up and one can solve effect of the back reaction
unambiguously.

\addtocounter{chapter}{1}
\setcounter{equation}{0}
\section*{\lromn
5. Model of Hawking evaporation
}

\vspace{0.5cm}
\hspace*{0.5cm}
The simplest spacetime that dynamically gives rise to the event horizon
would be half of the wormhole spacetime in a variant of two dimensional
dilaton gravity \cite{hy93-2}. This is an exact and the only sensible,
sensible as a model of gravitational collapse, classical solution
of a lagrangian field theory with the action,
\begin{equation}
S = \frac{1}{2\pi} \int\! d^{2} x \sqrt{-g}\;[\:e^{-2\varphi}
  \:( -R - 4\,\partial_{\mu}\varphi\,\partial^{\mu}\varphi + 4\lambda^{2})
+L^{(m)}\:]\,. \label{ourmodel}
\end{equation}
The field $\varphi $ here describes the dilaton.
The cosmological constant $\lambda^2$ sets a length scale in the theory.
When the matter part $L^{(m)}$ is taken to correspond to a local source,
with the stress tensor of
\begin{equation}
T_{++} =T_{--} =T_{+-} = -\,\frac{M}{4\pi}\,e^{\rho}\,\delta (x^1 )\,,
\label{source}
\end{equation}
the resulting metric in the conformal gauge,
\( \:
ds^{2}= -\,e^{2\rho}\,dx^{+}dx^{-}
\: \),
is the one given in Section \lromn2. This spacetime
is free from the curvature singularity in the usual sense despite the
presence of the source term $\: \propto \delta(x^{1})$.
Chung and Verlinde took a different model: the original CGHS model
\cite{cghs} which differs from our model Eq.\ref{ourmodel} in two
signs in the action.

The corresponding mirror trajectory in our model
is readily constructed as described
before. The question now is how to write a possible form of the action
leading to this mirror dynamics.
Although we feel that a more fundamental approach might
be contemplated, we shall be content here to give the mirror lagrangian and
discuss its equivalence to the wormhole spacetime.
In the next section we give an argument of how this particular model is
selected out in a class of mirror models.
With a minimum number of derivatives and the reparametrization invariance
in mind, the classical action we take is
\begin{equation}
S_{m} = m\, \int\, d\tau\,e^{-\lambda (z^{+}+z^{-})}\,
\sqrt{\dot{z}^{+}\dot{z}^{-}} \,,
\end{equation}
with the sign of $m$ chosen to match anticipated effects of the back reaction.
One might say that a non-trivial dynamics of the mirror particle is driven
by a varying mass factor
\( \:
e^{-\,2\lambda z^{0}}
\: \).

The classical equation that arises from this action is
\begin{equation}
\ddot{z}^{+} - \dot{z}^{+}\frac{\ddot{z}^{-}}{\dot{z}^{-}}
- 2\lambda \dot{z}^{+}(\dot{z}^{+}+\dot{z}^{-}) + 4\lambda \dot{z}^{+}
\dot{z}^{-} = 0 \,,
\end{equation}
and another equivalent equation with the replacement, \( \:
+ \leftrightarrow -
\: \).
The solution in the gauge of the proper time,
\( \:
\sqrt{\dot{z}^{+}\dot{z}^{-}} = 1
\:, \)
is given by
\begin{equation}
z^{+} = -\,\frac{1}{\lambda}\,\ln\sqrt{K}\cosh \lambda\tau
\,, \hspace{0.5cm} z^{-} = -\,\frac{1}{\lambda}\,\ln\sqrt{K}
\sinh \lambda\tau\,,
\end{equation}
with the parameter range \( \:
-\infty < \tau < 0
\:, \) ignoring inessential complications.
From this
\begin{equation}
e^{-\,2\lambda z^{+}} = e^{-\,2\lambda z^{-}} + K \,.
\end{equation}
This is equivalent to the one Eq.\ref{whole} in Section \lromn2, with
\( \:
K = \frac{M}{2\lambda }
\: \). This solution may also be derived directly in the gauge of
\( \:
\tau = z^{-}
\: \).
The mirror approaches a null line,
\( \:
z^{+} = x_{H} =
-\,\frac{1}{2\lambda} \ln\frac{M}{2\lambda}
\: \),
asymptotically at an exponential rate of \( \:
e^{-\,4\lambda t} \sim e^{-\,2\lambda z^{-}}
\: \).
From the expression of the mirror trajectory $z^{1}(z^{0})$ in
Eq.\ref{whtraject},
one may compute the recession velocity,
\begin{equation}
v = -\,\frac{Ke^{2\lambda t}}{\sqrt{4 + K^{2}e^{4\lambda t}}} \,,
\end{equation}
with $t = z^{0}$ the coordinate time.
With the precise form of the mirror trajectory known the semiclassical
flux is computed as
\begin{equation}
\langle \, T_{--} \, \rangle = \frac{\lambda^{2}}{12\pi }\,
\frac{1 + \frac{4\lambda }{M}\,e^{-\,2\lambda \sigma^{-}}}{(1 +
\frac{2\lambda }{M}\,e^{-\,2\lambda \sigma^{-}})^{2}} \,.
\end{equation}
As
\( \:
\sigma^{-} \rightarrow \infty \,,
\: \)
\( \:
\langle \, T_{--} \, \rangle \:\rightarrow \: \frac{\lambda^{2}}{12\pi } \,:
\: \)
the result Eq.\ref{aaflux} with $\kappa = 2\lambda $.

With the classical mirror dynamics set up, we may proceed to analysis of
the quantum back reaction caused by particle production. In this section
we shall choose the parameter $\gamma $ that appears in the back reaction
formula in Section \lromn4 to be vanishing.
With this choice the full equation for the mirror particle is
\begin{eqnarray}
&&
\frac{m}{2\sqrt{\dot{z}^{+}\dot{z}^{-}}}\,e^{-\lambda (z^{+}+z^{-})}\,
[\: \ddot{z}^{+} - \dot{z}^{+}\frac{\ddot{z}^{-}}{\dot{z}^{-}}
- 2\lambda \dot{z}^{+}(\dot{z}^{+}+\dot{z}^{-}) + 4\lambda \dot{z}^{+}
\dot{z}^{-} \:] = \nonumber \\
&&
-\,\frac{1}{4\pi}\,\frac{1}{\dot{z}^{-}}\,
[\:\frac{1}{6}\,\frac{\stackrel{...}{z}^{\,+}}{\dot{z}^{+}} - \frac{1}{4}\,
(\frac{\ddot{z}^{+}}{\dot{z}^{+}})^{2} -
\frac{1}{6}\,\frac{\stackrel{...}{z}^{\,-}}{\dot{z}^{-}} +
\frac{1}{4}\,(\frac{\ddot{z}^{-}}{\dot{z}^{-}})^{2} \:]
 \,,
\end{eqnarray}
and another equivalent equation with the replacement, \( \:
+ \leftrightarrow -
\: \).
With a gauge choice the dynamical equation is simplified. For instance in
the gauge of \( \:
\tau = z^{-}
\: \), the independent equation is written as
\begin{eqnarray}
\frac{m}{2}\,[\:\ddot{z}^{+} - 2\lambda \dot{z}^{+}(\dot{z}^{+} -1)\:] =
-\,\frac{1}{24\pi}\, e^{\lambda (z^{+}+\tau)}\,\dot{z}^{+}\,
\frac{d}{d\tau}[\:\ddot{z}^{+}(\dot{z}^{+})^{-3/2}\:]\,.
\end{eqnarray}

As a check of the effect of back reaction we make an adiabatic approximation
assuming that the mass parameter $K$ is slowly varying. With higher derivatives
ignored,
\begin{equation}
\frac{m}{2}\,\dot{K} \sim  -\,\frac{\lambda^{2}}{12\pi}\,
\frac{Ke^{2\lambda \tau}\,(Ke^{2\lambda \tau}+2)}
{(Ke^{2\lambda \tau}+1)^{2}} \,,
\end{equation}
with $\tau = z^{-}$.
This equation precisely describes how the collapsed mass $M = 2\lambda K$
decreases due to Hawking radiation; the right hand side of this equation
approaches a constant value
\( \:
-\,\frac{\lambda^{2}}{12\pi }
\: \), asymptotically as
\( \:
\tau \: \rightarrow \: \infty \,,
\: \) as anticipated.

In order to further analyze the behavior of the mirror point it is convenient
to write the differential equation in terms of the mirror velocity $v$.
Using the coordinate time $t = z^{0}$ as the parameter for the trajectory,
the dynamical mirror equation reads as
\begin{eqnarray}
m(1 - v^{2})^{1/2}\,
e^{-\,2\lambda t}\,[\:v' - 2\lambda v\,(1-v^{2})\:]
= -\,\frac{1}{12\pi }\,[\:v'' + 3v'^{2}\,\frac{v}{1-v^{2}}\:] \,, \label{meq}
\end{eqnarray}
with the prime indicating the $z^{0}-$derivative.
It should be obvious that the velocity $v$ is restricted;
\( \:
|v| \leq 1
\: \).

An immediate general result follows from this equation. The mirror particle
at intermediate times cannot stop accelerating even momentarily;
\( \:
v' \neq 0
\: \). To prove this, suppose that $v' = 0$ at some finite time.
With \( \:
|v| < 1
\: \) the mirror equation \ref{meq} would imply that \( \:
vv'' > 0
\: \) there. With the initial condition $v<0$ it means that the behavior of $v$
is convex near its extremum; a result incompatible with the initial decrease
of $v$: $\:v' < 0$. Thus one may assert that always \( \:
v <0  \: \mbox{and} \: v' <0
\: \) at any intermediate times.

One may analyze the end point behavior using the differential equation
for the velocity. First, let us enumerate the general form of the mirror
motion in which the back reaction ceases to operate.
Setting the right hand side of Eq.\ref{meq}  to vanish,
one finds that either \( \:
v = v_{0} \:
\: \) with a constant $v_{0}$ of
\( \:
| v_{0} | < 1 \,,
\: \)
or
\begin{eqnarray}
v = -\,\frac{At}{\sqrt{1+A^{2}t^{2}}} \: \equiv f_{0}(t)\: \,.
\end{eqnarray}
In the second case
\begin{eqnarray}
z^{+} = t - \frac{1}{A}\,\sqrt{1+A^{2}t^{2}} + \mbox{const.} \,,
\end{eqnarray}
or equivalently
\( \:
z^{+} = p(z^{-}) = B - \frac{A^{-2}}{z^{-} + B} \,.
\: \)
The second possibility is known as the motion of uniform acceleration
in the literature \cite{davful77},
because in the instantaneous rest frame the acceleration
$\alpha $ defined by Eq.\ref{acceleration} is constant;
\begin{equation}
\alpha  = \frac{v'}{(1-v^{2})^{3/2}}
= -\,A \,.
\end{equation}

By considering the full equation \ref{meq} one then proves
that the approach towards a constant $v_{0} \neq -1$ is
impossible; the first possibility is rejected.
First, the final velocity $v_{0}$ must be negative to conform
to the initial negative acceleration together with
the impossibility of $v' =0$.
Then the perturbation analysis indicates that
\begin{equation}
v \sim v_{0} + \delta v \,, \hspace{0.2cm} \mbox{with} \hspace{0.2cm}
\delta v = \frac{6\pi m}{\lambda }\,v_{0}\,(1-v_{0}^{2})^{3/2}\,
e^{-\,2\lambda t} < 0 \,.
\end{equation}
Thus starting from the initial $v = 0$, the mirror velocity approaches
the final value $v_{0} < 0$ from below.
This would however be incompatible with $v' \neq  0$ at any finite time.
We thus singled out the
possible mirror motion allowed for the end point behavior: recession with
a uniform acceleration.

One can work out the approach to the asymptotic mirror motion by solving the
perturbation around the end point behavior $v = f_{0}(t)$.
With \( \:
v = f_{0}(t) + \delta v
\: \), the asymptotic perturbation equation reads with
\( \:
c = \frac{1}{12\pi m}
\: \)
as
\begin{equation}
-\,c\,\delta v'' - \frac{6c}{t^{2}}\,(t\,\delta v' + \delta v) =
\frac{2\lambda }{A^{3}t^{3}}\,e^{-\,2\lambda t} \,,
\end{equation}
for \( \:
t \: \rightarrow \: \infty
\: \).
Thus the deviation of the mirror trajectory from the end point formula
is given by
\begin{eqnarray}
\delta v \sim -\,\frac{1}{2c\lambda A^{3}}\,t^{-3}\,e^{-\,2\lambda t} \,,
\end{eqnarray}
hence
\begin{eqnarray}
z^{+} \sim \mbox{const.} - \frac{1}{A^{2}z^{-}} + \frac{2}{c\lambda^{2}A^{3}}\,
(z^{-})^{-3}\,e^{-\,\lambda z^{-}} \,.
\end{eqnarray}
The approach to the end point behavior is thus exponentially rapid.
For instance,
\begin{eqnarray}
\langle\, T_{--}\, \rangle  = O\,[\: \frac{1}{\sigma^{-}}
\,e^{-\,\lambda \sigma^{-}}\:] > 0 \,.
\end{eqnarray}

Dependence on the mirror mass $m$ is not arbitrary. This parameter in
the mirror equation is changed by shifting the time coordinate $z^{0}$.
The time shift also changes the classical mass parameter $M$,
but the combination
of their product $mM$ is invariant under the time translation. Thus
ignoring the absolute position of the time origin, all physical results
must depend on $m$ in the combination of $mM$.

Although our analytic method is adequate to establish the end point
behavior, we did for completeness a sample numerical integration to
the differential equation
to the mirror motion. A typical behavior of the asymptotic flux
\( \:
\langle \, T_{--} \, \rangle
\: \)
that exhibits the semiclassical
Hawking picture at intermediate times and the one that does not are
shown in Fig. 2 and Fig. 3 for various models of different $\gamma $
($\gamma \neq 0 $ case to be discussed later)
and different classical mass parameter $M$.
Difference in Fig. 2 and Fig. 3 lies in the strength of the back reaction
measured by the magnitude of $mM$, differing by a factor $10^{8}$ in the
two figures. For
\( \:
mM = O\,[\,\lambda^{2} \,] \,,
\: \)
there is no resemblance to the Hawking picture since quantum effects
become sizable even before any precursory symptom of event horizon
formation occurs.
In Fig. 4 is shown the behavior of the mirror trajectory for
the same set of parameters as in Fig. 3.

The end point behavior of the mirror motion is quite unique in many respects.
With the precise form of the trajectory function at the end point given by
\begin{eqnarray}
p(\sigma^{-}) = B - \frac{A^{-2}}{\sigma^{-} + B}
\hspace{0.5cm} \mbox{for} \hspace{0.2cm} \sigma^{-} > -B \,,
\label{ltrajectory}
\end{eqnarray}
with a correction of order
\( \:
(\sigma^{-})^{-3}\,e^{-\,\lambda \sigma^{-}} \,,
\: \)
the correlation exactly coincides with the free field value
of the fixed boundary case,
\begin{eqnarray}
\langle\, \partial_{-}f(\sigma)\partial_{-}f(\sigma')\, \rangle\,_{\rm{in}}
= -\,\frac{1}{4\pi }\,\frac{1}{[\:\sigma^{-}-\sigma'^{-}-i\epsilon \:]^{2}}
\,, \label{fcorrelation}
\end{eqnarray}
if one ignores exponentially small terms.
Thus, combined  with the vanishing outgoing flux of radiation,
\( \:
\langle\, T_{--}\, \rangle = 0
\: \), there is no sign of loss of quantum coherence in the end point.
As will be shown in the following section,
this would not occur for the other choices
of the parameter $\gamma $ in the back reaction formula.

Although no energy flux is detected to an asymptotic observer in the
end, the Bogoliubov coefficients do not vanish
for this uniform acceleration \cite{davful77}.
Taking the simple form of the late time trajectory Eq.\ref{ltrajectory},
one finds that
\begin{eqnarray}
\tilde{\beta}_{\omega'\omega}^{R} &=&
\frac{1}{2\sqrt{\omega \omega'}}\, \int_{-B}^{\infty }\,d\sigma\,
[\:\omega - \omega' p'(\sigma )\:]\,
e^{i(\omega \sigma + \omega'p)} \nonumber \\
&=&
\frac{1}{2A}\,e^{iB(\omega' - \omega )}\,\int_{-\infty }^{\infty }\,du\,
[\: e^{-\,i(2/A)\sqrt{\omega'\omega}\,\sinh u - u} -
e^{-\,i(2/A)\sqrt{\omega'\omega}\,\sinh u + u} \:] \nonumber \\
&=&
i\,\frac{1}{2A}\,e^{iB(\omega' - \omega)}\:K_{1}(2\sqrt{\omega'\omega }/A)
\,, \\
\tilde{\alpha}_{\omega'\omega}^{R} &=&
-\,\frac{1}{2A}\,e^{-\,iB(\omega' + \omega)}\:
K_{1}(i\,2\sqrt{\omega'\omega }/A)
\,,
\end{eqnarray}
hence
\( \:
\langle\, a_{\omega }^{R \dag }a_{\omega'}^{R}\, \rangle\,_{\rm{in}}
\neq 0 \,,
\: \)
and
\( \:
\langle\, a_{\omega }^{R}a_{\omega'}^{R}\, \rangle\,_{\rm{in}} \neq 0 \,.
\: \)

The fact that incoming radiation incident on the mirror at very late
times never gets reflected clearly indicates that a remnant is left
behind. The end point behavior of uniform mirror acceleration is
something not anticipated when we first classified the end point
behavior of the mirror trajectory at the end of the preceding section.
What precisely is left behind is not known at present, and one presumably
needs an approach somewhat different from that taken so far,
in order to gain a deeper insight into the remnant.

We wish, however, to point out that the left-right correlation such as
observed in the case of event horizon formation is absent here.
From
\begin{equation}
\langle\, \partial_{-}f(\sigma )\,\partial_{+}f(\sigma' )\, \rangle\,
_{\rm{in}} =
-\,\frac{1}{4\pi }\,\frac{p'(\sigma^{-})}{[\:p(\sigma^{-}) - \sigma'^{+}
- i\epsilon \:]^{2}} \,,
\end{equation}
one finds for $\sigma $ in the thermal stage and $\sigma'$ in the
end point stage that
\begin{equation}
\langle\, \partial_{-}f(\sigma )\,\partial_{+}f(\sigma' )\, \rangle\,
_{\rm{in}} \sim
-\,\frac{1}{4\pi }\,\frac{2\lambda/M}{(\sigma'^{+} - x_{H})^{2}}
\,e^{-\,2\lambda  \sigma^{-}} \,.
\end{equation}
The limiting value that results as $\sigma'^{+}$ approaches the null
line at $B$ is vanishingly smaller compared to the thermal case
Eq.\ref{thlimit}, the ratio being
\( \:
\left( M\,(B - x_{H})\, \right)^{-2}\,e^{-\,4\lambda  \sigma^{-}}
\: \).
Thus there is no left-right correlation remaining.

In retrospect, one could have enumerated beforehand
all possible forms of the mirror trajectory leading to the "free" field
correlation. Indeed by solving the functional equation for $p(\sigma )$
\begin{equation}
\frac{p'(\sigma )\,p'(\sigma')}{[\, p(\sigma ) - p(\sigma') \,]^{2}} =
\frac{1}{(\, \sigma - \sigma' \,)^{2}} \,,
\end{equation}
that expresses agreement of the two point correlation
\( \:
\langle\, \partial_{-}f(\sigma )\,\partial_{-}f(\sigma' )\, \rangle\,
_{\rm{in}}
\: \)
with the case of the fixed boundary,
one finds the most general solution of the form
\begin{equation}
p(\sigma ) = \frac{A\sigma + B}{C\sigma + D} \,,
\end{equation}
with $A-D$ integration constants. The trajectory thus found essentially
fall into two classes: a constant velocity case with $C = 0$ and
a uniform acceleration case with $C \neq 0$.
For $C \neq 0$ the cross correlation
\( \:
\langle\, \partial_{-}f(\sigma )\,\partial_{+}f(\sigma' )\, \rangle\,
_{{\rm in}}
\: \)
does not agree with the fixed boundary case.
In our model of the back
reaction only one class of the uniform acceleration is realized.

It is perhaps most interesting to closely examine how the thermal stage
of Hawking evaporation decays to finally recover the quantum mechanical
correlation. Suppose that the semiclassical thermal stage lasts for a
long time interval
\( \:
\Delta \sigma \gg 1/2\lambda
\: \).
Take as an initial fiducial time
\( \:
\sigma^{-} = \sigma_{0}
\: \)
during the thermal period. The observable correlation
\( \:
\langle\, \partial_{-}f(\sigma_{0})\,\partial_{-}f(\sigma )\, \rangle\,
_{\rm{in}}
\: \)
seen by observers at the right null infinity $I_{R}^{+}$ behaves in the
following way as the time $ \sigma^{-} = \sigma $ increases.
First in the shortest time limit of
\( \:
\sigma \rightarrow \sigma_{0}
\: \)
the "vacuum" correlation of
\( \:
-\,\frac{1}{4\pi }\,(\sigma_{0} - \sigma - i\epsilon )^{-2}
\: \)
is present. This is followed by a long period of thermal correlation
of short range, behaving like
\( \:
-\,\frac{\lambda^{2}}{\pi }\,e^{-2\lambda |\sigma_{0} - \sigma |}
\: \).
However this thermal correlation does not last forever, and for
\( \:
|\sigma_{0} - \sigma | \gg \Delta \sigma
\: \),
\begin{equation}
\langle\, \partial_{-}f(\sigma_{0})\,\partial_{-}f(\sigma )\, \rangle\,
_{\rm{in}}
\: \rightarrow \:
-\,\frac{1}{4\pi }\,\frac{2\lambda A^{-2}/M}{(B - x_{H})^{2}}\,
\frac{e^{-\,2\lambda  \sigma^{0}}}{(\sigma + B)^{2}} \,.
\end{equation}
Namely, there exists a long tail of correlation decreasing only by a power
$\sigma^{-2}$ between the thermal and the final vacuous stage.
Correlation more directly measurable is manifested in \cite{carlitz}
\begin{equation}
\langle\, T_{--}(\sigma_{0})\,T_{--}(\sigma )\, \rangle\,_{\rm{in}} -
\langle\, T_{--}(\sigma_{0})\, \rangle\,\langle\, T_{--}(\sigma )\, \rangle =
\langle\, \partial_{-}f(\sigma_{0})\,\partial_{-}f(\sigma )\, \rangle\,
_{\rm{in}}^{2} \,.
\end{equation}
This long term correlation is stronger at an earlier stage of the
thermal period because of the factor
\( \:
e^{-2\lambda \sigma_{0}}
\: \).
We stress that the system viewed at the right null infinity $I_{R}^{+}$
obeys the law of quantum mechanics, but it is an unusual quantum system
characterized by a long range correlation in spite of the local
thermal behavior at intermediate times.

The simple end point behavior of the correlation function is related to
a special Lorentz invariance of the final mirror trajectory. By appropriately
shifting the origin of coordinates one may write the mirror trajectory
in the new coordinates $\tilde{\sigma}^{\pm }$ as
\begin{equation}
\tilde{\sigma }^{+}\,\tilde{\sigma }^{-} = -\,A^{-2} \,.
\end{equation}
This trajectory is invariant under a Lorentz boost,
\begin{equation}
\tilde{\sigma }^{+} \rightarrow e^{\mu }\,\tilde{\sigma }^{+}
\,, \hspace{0.5cm}
\tilde{\sigma }^{-} \rightarrow e^{-\,\mu }\,\tilde{\sigma }^{-} \,,
\end{equation}
with $\mu $ a real constant.
Indeed, the correlation Eq.\ref{fcorrelation}
respects this Lorentz covariance, as it should be.
The entire two point correlation may be recast into a manifestly
Lorentz invariant form;
\begin{equation}
\langle\, f(\tilde{\sigma })f(\tilde{\sigma' })\, \rangle\,_{\rm{in}} =
-\,\frac{1}{4\pi }\,\ln\,\frac{A^{-2}(\Delta\tilde{\sigma }^{+} - i\epsilon )\,
(\Delta\tilde{\sigma }^{-} - i\epsilon )}
{(\tilde{\sigma}^{+}\tilde{\sigma}'^{-} + A^{-2} - i\epsilon )\,
(-\,\tilde{\sigma}'^{+}\tilde{\sigma}^{-} - A^{-2} - i\epsilon )} \,,
\end{equation}
with
\( \:
\Delta\tilde{\sigma }^{\pm } = \tilde{\sigma }^{\pm } - \tilde{\sigma }'^{\pm }
\: \).

There is even a larger dynamical symmetry that manifests as a global
symmetry such as the Lorentz invariance above when restricted to a limited
spacetime region. This is a spacetime
dependent symmetry that locally leaves the mirror trajectory invariant,
and at initial times manifests as the time translation symmetry because
the mirror is initially at rest. Presumably a deeper understanding of
our results is gained from this symmetry viewpoint. We wish to come back
to this point in the future.

One may reconstruct the spacetime metric from the mirror trajectory as
described in Section \lromn2.
In the original coordinate $x^{\pm }$ has a semi-infinite
range;
\( \:
x^{-} < \frac{C}{A}
\: \), and the metric at the end point is given by
\begin{eqnarray}
ds^{2} \sim -\,\frac{dx^{+}dx^{-}}{(\,C-Ax^{-}\,)^{2}} \,.
\end{eqnarray}
The spacetime is flat except at the origin.
The finite curvature at the origin is computed from the formula,
\begin{equation}
R = 8e^{-2\rho }\,\partial_{+}\partial_{-}\rho
= -\,4p''(\sigma^{-})\,\frac{\partial \sigma ^{-}}{\partial x^{-}}
\,\delta (x^{1}) \,.
\end{equation}
The end point behavior thus found at the origin is
\begin{equation}
R \sim  8CA\,(1-\frac{A}{C}\,x^{-})\,\delta (x^{1}) \,.
\end{equation}
The curvature looks gradually diminishing to an asymptotic observer
since to him the relevant coordinate is
\begin{equation}
\sigma^{-} \sim (A)^{-1}\frac{1}{C - Ax^{-}} \,.
\end{equation}

\addtocounter{chapter}{1}
\setcounter{equation}{0}
\section*{\lromn
6. Extension to other models
}

\vspace{0.5cm}
\hspace*{0.5cm}
In the preceding section we discussed a particular model of mirror dynamics
and found some interesting results on the end point behavior of the Hawking
evaporation. In this section we briefly mention generalization of models.
Extension is directed in two ways : first to consider the cases of modified
back reaction in which $\gamma \neq 0$ ,
and second to generalize the classical action of mirror dynamics.

Let us first write down the mirror equation in terms of the velocity $v$
in the generalized cases in which
the back reaction parameter $\gamma \neq 0$;
\begin{eqnarray}
m\,e^{-\,2\lambda t}\,[\:v' - 2\lambda v\,(1-v^{2})\:]
= -\,\frac{1}{12\pi }\,(1 - v^{2})^{-1/2}\,
[\:v'' + v'^{2}\,\frac{3v - 2\gamma }{1-v^{2}}\:] \,.
\end{eqnarray}
As in the previous case of $\gamma  = 0$ one verifies that the velocity
monotonically decreases as the time $t$ increases.
The end point behavior is derived by setting
the right hand side to vanish in this equation. With
\( \:
\gamma > -\,\frac{1}{2}
\: \),
this leads to the following end point behavior;
\begin{eqnarray}
\nu t &\sim& \int_{v}\,dv\,(1-v)^{-\frac{3}{2}+\gamma }\,(1+v)^{-\frac{3}{2}
-\gamma } \nonumber \\
&= &
\frac{1}{2}\,[\: \frac{1}{1+2\gamma }\,(\frac{1-v}{1+v})^{\frac{1}{2}+\gamma }
- \frac{1}{1-2\gamma }\,(\frac{1+v}{1-v})^{\frac{1}{2}-\gamma } \:]
+ \mbox{const.} \,.
\end{eqnarray}
The special case of
\( \:
\gamma = -\,\frac{1}{2}
\: \)
must be treated separately. Thus for
\( \:
\gamma \neq -\,\frac{1}{2} \,,
\: \)
as \( \:
t \rightarrow \infty \,,
\: \)
\begin{equation}
v \: \rightarrow \:  -1 + \tilde{\nu }\,t^{-2/(1+2\gamma )} \,,
\end{equation}
hence
\begin{equation}
z^{+} \:\rightarrow  \:
\mbox{const.} - \frac{1+2\gamma }{1-2\gamma }\,\tilde{\nu }\,
t^{-\,\frac{1-2\gamma }{1+2\gamma }} \,,
\end{equation}
with
\( \:
\tilde{\nu } = [\:(1 + 2\gamma )\,2^{\frac{1}{2}-\gamma }\,\nu  \:]
^{-\,2/(1+\gamma )} \,.
\: \)
The stress tensor behaves like
\begin{equation}
\langle\, T_{--}\, \rangle \sim  -\,\frac{1}{24\pi }
\,\frac{\gamma }{(1+2\gamma )^{2}}
\,\frac{1}{t^{2}} \,,
\end{equation}
and a complicated correlation, neither agreeing with that of the fixed
boundary case nor with that of the thermal case, remains.
The end point behavior of the spacetime metric is given by
\begin{equation}
ds^{2} \: \propto \:
-\,(x_{0} - x^{-})^{\frac{2}{2\gamma - 1}}\,dx^{+}dx^{-} \,.
\end{equation}

In the special case of
\( \:
\gamma = -\,\frac{1}{2}
\: \)
one finds that
\begin{eqnarray}
\nu t \sim -\,\frac{1}{4}\,[\:\ln \frac{1+v}{1-v} + \frac{1+v}{1-v}\:]
\: \rightarrow \:
-\,\frac{1}{4}\,\ln (1+v) \,,
\end{eqnarray}
hence
\begin{eqnarray}
z^{+} \sim x_{0} - C_{0}\,e^{-\,2\nu z^{-}} \,,
\end{eqnarray}
as $t$ or
\( \:
z^{-} \rightarrow \infty \,.
\: \)
Although the classical behavior of the mirror motion is modified, the end
point still has an approach to a null line with an exponential rate.
Thus the event horizon remains in this special case, and there is a finite
modified flux of Hawking radiation
\begin{eqnarray}
\langle\, T_{--}\, \rangle \sim \frac{\nu^{2}}{12\pi } \,.
\end{eqnarray}

Some examples of numerical integration for models of $\gamma \neq 0$
are shown in Fig. 2 to Fig. 4. One can observe in these models behaviors
entirely different from the model of $\gamma = 0$.
The asymptotic flux
\( \:
\langle \, T_{--} \, \rangle
\: \)
is the best quantity to distinguish various models. Models of
$\gamma = 0$ and $\gamma = -\,\frac{1}{2}$ stand out as anticipated.

Summarizing this class of extension, one may conclude that only the case of
\( \:
\gamma = 0
\: \)
considered previously
yields an entirely unitary evolution seen at the right null infinity,
and that there is even a situation in which Hawking radiation never stops.

Our classical mirror model is based on a well defined dilaton theory in
two dimensions. However it would be interesting to pursue the mirror model
in its own right. In this respect one might attempt to classify a set of
mirror models whose actions give rise to a reasonable classical behavior
appropriate for our discussion of the gravitational collapse. With a minimum
set of assumptions what is achieved is very much limited, nevertheless
we shall describe our finding. Consider a class of reparametrization
invariant mirror action of the form,
\begin{equation}
S_{m} = \int\, d\tau\,l(z^{+}\,,z^{-})\,
\sqrt{\dot{z}^{+}\dot{z}^{-}} \,.
\end{equation}
Equation of motion follows from the variational principle;
\begin{equation}
[\: \frac{1}{4}\,(\frac{\ddot{z}^{+}}{\dot{z}^{+}} -
\frac{\ddot{z}^{-}}{\dot{z}^{-}})\,l
+ \frac{1}{2}\,(\,\dot{z}^{+}\,\partial_{+}l
- \dot{z}^{-}\,\partial_{-}l\,)\:]\,\sqrt{\frac{\dot{z}^{+}}{\dot{z}^{-}}} =
F_{-} \,,
\end{equation}
and a similar equation for the replacement,
\( \:
+ \leftrightarrow -
\: \).

We may derive constraint on the form of allowed $l(z^{+}\,,z^{-})$ by
imposing that
the mirror trajectory corresponding to the event horizon formation
should be a consistent solution to the equation of mirror motion.
For this purpose we first write the mirror equation
without the back reaction term $F_{-}$ in terms of the velocity,
\begin{equation}
\frac{v'}{1 - v^{2}}\,l + [\: (1+v)\,\partial_{+}l - (1-v)\,\partial_{-}l\:]
= 0 \,.
\end{equation}
We further make a simplifying ansatz for $z^{\pm }$ dependence of
the unknown function $l$:
\( \:
l(z^{+}\,,z^{-}) = l(z^{+} + \eta z^{-})
\: \)
with $\eta $ a constant,
and put in the asymptotic form of the solution corresponding to the formation
of event horizon,
\( \:
v \sim -1 + 2\kappa \xi \,e^{-\,2\kappa t}
\: \).
This consideration determines the allowed form of $l(z^{+}\,,z^{-})$ in the
asymptotic region,
\begin{equation}
l(z^{+}\,,z^{-}) = m\,e^{-\,\frac{\kappa }{2\eta }\,(z^{+} +
\eta z^{-})}\,.
\end{equation}
Back to the mirror equation with this given,
\begin{eqnarray}
\frac{2v'}{1-v^{2}} - \frac{\kappa }{\eta }\,[\:(1-\eta ) + (1+\eta )\,
v\:] = 0 \,,
\end{eqnarray}
one has an improved solution in the late time;
\begin{eqnarray}
\frac{[\:1-\eta + (1+\eta )\,v\:]^{1+\eta }}{(1-v)^{\eta }\,(1+v)} =
C^{2}e^{2\kappa t} \,.
\end{eqnarray}
If we trust the form of $l(z^{+}\,,z^{-})$ in the entire spacetime region,
not necessarily restricted to late times, we may
reject all solutions except in the case of
\( \:
\eta = 1
\: \), because of a bad past behavior of the velocity; the mirror does not
start from the state at rest in the cases of $\eta \neq 1$.
This argument in particular rules out another interesting possibility of
\( \:
\eta = -1
\: \), in which case
\( \:
l(z^{+}\,,z^{-}) = m\,e^{\kappa z^{1}}
\: \): a space dependent mass term.

Arguments thus far presented, although by no means compelling, suggest a
priviledged status of our classical mirror model and our treatment of
the quantum back reaction in the previous section.
It would be extremely interesting if one could
justify this mirror model from a systematic truncation of the four dimensional
theory.

\newpage

\appendix
\begin{center}
\begin{Large}
{\bf Appendix}
\end{Large}
\end{center}

\vspace{0.5cm}
\addtocounter{chapter}{1}
\setcounter{equation}{0}
\section*
{A.
General properties of the Bogoliubov coefficients}

\vspace{0.5cm}
\hspace*{0.5cm}
Let us first recall the in-mode and the out-mode decomposition in our problem,
\begin{eqnarray}
f(\sigma)
&=& \frac{1}{2\pi}\, \int_{0}^{\infty}\,d\omega\,
(f_{\omega}a_{\omega}+f_{\omega}^{*}a_{\omega}^{\dag}) \nonumber \\
&=& \frac{1}{2\pi}\, \sum_{I=R,L}\,\int_{0}^{\infty}\,d\omega\,
(f_{\omega}^{I}\,a_{\omega}^{I}+f_{\omega}^{I *}\,a_{\omega}^{I
\dag}) \,,
\end{eqnarray}
with
\begin{equation}
a_{\omega} =
i\,\int_{-\infty}^{\infty}\,d\sigma^{+}\,f^{*}_{\omega}\,\tilde{\partial}_{+}
f(\sigma) \,, \hspace{0.5cm}
a_{\omega}^{I} = i\,\int_{I^{+}_{L}\cup I^{+}_{R}}
\,d\sigma^{\mu}\,f^{I *}_{\omega}\,\tilde{\partial}_{\mu} f(\sigma) \,,
\end{equation}
where the Klein-Gordon inner product is defined by
\begin{equation}
i\,\int\,d\sigma^{\mu }\,
f_{1}^{*}\tilde{\partial}_{\mu } f_{2} \equiv
i\,\int\,d\sigma^{\mu }\,[\:f_{1}^{*}\partial_{\mu }f_{2} -
f_{2}\partial_{\mu }f_{1}^{*}\:]
\,. \label{kgproducta}
\end{equation}
Mode mixing equations are then
\begin{eqnarray}
a_{\omega}
&=& \frac{1}{2\pi}\, \sum_{I=R,L}\,\int_{0}^{\infty}\,d\omega'\,
(\alpha_{\omega\omega'}^{I}\,a_{\omega'}^{I}+
\beta_{\omega\omega'}^{I}\,a_{\omega'}^{I \dag }) \,, \\
\alpha_{\omega\omega'}^{I} &=&
i\,\int_{-\infty}^{\infty}\,d\sigma^{+}\,f^{*}_{\omega}\,\tilde{\partial}_{+}
f_{\omega'}^{I}
\,, \hspace{0.5cm}
\beta_{\omega\omega'}^{I} =
i\,\int_{-\infty}^{\infty}\,d\sigma^{+}\,f^{*}_{\omega}\,\tilde{\partial}_{+}
f_{\omega'}^{I *} \,, \\
a_{\omega}^{I}
&=& \frac{1}{2\pi}\,\int_{0}^{\infty}\,d\omega'\,
(\tilde{\alpha}_{\omega'\omega}^{I}\,a_{\omega'}+
\tilde{\beta}_{\omega'\omega}^{I}\,a_{\omega'}^{\dag }) \,, \\
\tilde{\alpha}_{\omega'\omega}^{I} &=&
i\,\int_{I^{+}_{L}\cup I^{+}_{R}}\,d\sigma^{\mu }
\,f^{I *}_{\omega}\,\tilde{\partial}_{\mu }f_{\omega'}
\,, \hspace{0.5cm}
\tilde{\beta}_{\omega'\omega}^{I} =
i\,\int_{I^{+}_{L}\cup I^{+}_{R}}\,d\sigma^{\mu }
\,f^{I *}_{\omega}\,\tilde{\partial}_{\mu }f_{\omega'}^{*} \,.
\end{eqnarray}

Exact symmetry relations for the Bogoliubov coefficients follow by noting
the invariance of the Klein-Gordon inner product Eq.\ref{kgproducta};
\begin{eqnarray}
\tilde{\alpha}_{\omega'\omega}^{R} = \alpha_{\omega'\omega}^{R *}
\,, \hspace{0.5cm}
\tilde{\alpha}_{\omega'\omega}^{L} = \alpha_{\omega'\omega}^{L *}
\,, \hspace{0.5cm}
\tilde{\beta}_{\omega'\omega}^{L} = -\,\beta_{\omega'\omega}^{L}
\,, \hspace{0.5cm}
\tilde{\beta}_{\omega'\omega}^{R} = -\,\beta_{\omega'\omega}^{R}
\,.
\end{eqnarray}
Furthermore, when propagating mode functions of the form
\( \:
e^{\pm i\omega a(\sigma)}
\: \)
are adopted with \( \:
a(\sigma)
\: \) a real function, one finds an additional relation,
\begin{equation}
\alpha_{\omega'\omega}^{I} = \beta_{\omega' -\omega}^{I} \,.
\end{equation}
These formulas may be used to simplify some general discussion.

We next write the completeness relation for the Bogoliubov coefficients.
Using the in-mode and the out-mode commutation relations,
\begin{equation}
[\:a_{\omega}\,,\:a_{\omega'}^{\dag}\:] = 2\pi\,\delta(\omega-\omega') \,,
\hspace{0.5cm}
[\:a_{\omega}^{I}\,, \: a_{\omega'}^{J \dag}\:] = 2\pi\,
\delta(\omega-\omega')\,\delta_{I,J} \,,
\end{equation}
and
\( \:
[\: a_{\omega } \,, \: a_{\omega' } \:] = 0 \,, \hspace{0.2cm}
[\: a_{\omega }^{I} \,, \: a_{\omega' }^{J} \:] = 0 \,,
\: \)
one derives
\begin{eqnarray}
&&
\frac{1}{2\pi}\,\sum_{I=R,L}\, \int_{0}^{\infty}\,d\omega''\,
[\:\alpha_{\omega\omega''}^{I}\,\alpha_{\omega'\omega''}^{I *}-
\beta_{\omega\omega''}^{I}\,\beta_{\omega'\omega''}^{I *}\:]
= 2\pi\delta(\omega-\omega') \,,
\\ &&
\frac{1}{2\pi}\,\sum_{I=R,L}\, \int_{0}^{\infty}\,d\omega''\,
[\:\alpha_{\omega\omega''}^{I}\,\beta_{\omega'\omega''}^{I}-
\beta_{\omega\omega''}^{I}\,\alpha_{\omega'\omega''}^{I}\:]
= 0 \,,
\end{eqnarray}
by recalling the definition of the Bogoliubov coefficients.
Similarly
\begin{eqnarray}
&&
\frac{1}{2\pi}\, \int_{0}^{\infty}\,d\omega^{''}\,
[\:\tilde{\alpha}_{\omega''\omega}^{I}\,\tilde{\alpha}_{\omega''\omega'}^{J *}-
\tilde{\beta}_{\omega''\omega}^{I}\,\tilde{\beta}_{\omega''\omega'}^{J *}\:]
= 2\pi\delta(\omega-\omega')\,\delta_{I,J} \,, \label{completeness}
\\ &&
\frac{1}{2\pi}\, \int_{0}^{\infty}\,d\omega^{''}\,
[\:\tilde{\alpha}_{\omega''\omega}^{I}\,\tilde{\beta}_{\omega''\omega'}^{J}-
\tilde{\beta}_{\omega''\omega}^{I}\,\tilde{\alpha}_{\omega''\omega'}^{J}\:]
= 0 \,.
\end{eqnarray}

\vspace{0.5cm}
\addtocounter{chapter}{1}
\setcounter{equation}{0}
\section*
{
B.
Explicit formulas and some relations of \\
Bogoliubov coefficients for our modes}

\vspace{0.5cm}
\hspace*{0.5cm}
The mirror model most extensively discussed in the text is defined
by the trajectory function  and the in-mode function of the form
\begin{eqnarray}
&& p(\sigma^{-}) = \sigma^{-}-
\frac{1}{2\lambda} \ln (1+\frac{M}{2\lambda} e^{2\lambda \sigma^- } )\,, \\
&& f_{\omega} = \frac{1}{\sqrt{2\omega}}\,[\,
e^{-i\omega\sigma^{+}} - e^{-i\omega\sigma^{-}}
\,(1+\frac{M}{2\lambda} e^{2\lambda \sigma^- }\,)^{i\frac{\omega}{2\lambda}}\:]
\,.
\end{eqnarray}
The right moving out-modes are given, with the inversion of
\( \: \sigma^{+} = p(\sigma^{-})  \:
\rightarrow \: \sigma^{-} = q(\sigma^{+}) \,,\:\)
\begin{equation}
q(\sigma^{+}) =
-\,\frac{1}{2\lambda} \ln (e^{-2\lambda\sigma^+} - \frac{M}{2\lambda})\,,
\end{equation}
by
\begin{eqnarray}
f^{R}_{\omega}
&=&
\frac{1}{\sqrt{2\omega}}\, [\:
-\,e^{-i\omega\sigma^{-}}+e^{-i\omega q(\sigma^{+})}\,
\theta(x_{H} - \sigma^{+})\:] \, \nonumber \\
&=&
\frac{1}{\sqrt{2\omega}}\, [\: -\,e^{-i\omega\sigma^{-}}
+ (e^{-2\lambda\sigma^{+}}-\frac{M}{2\lambda})^{i\frac{\omega}{2\lambda}}\:
\theta(x_{H} - \sigma^{+})\:] \,,
\end{eqnarray}
with \( \:
x_H =-\,\frac{1}{2\lambda} \ln\frac{M}{2\lambda}\,.
\: \)
By computing the overlap of the mode functions, one finds that
\begin{eqnarray}
&&
\tilde{\alpha }_{\omega'\omega}^{R *} = \alpha_{\omega'\omega}^{R} =
\frac{1}{2\lambda}\,\sqrt{\frac{\omega}{\omega'}}\,
(\frac{M}{2\lambda})^{i\frac{\omega-\omega'}{2\lambda}}\,
B(i\,\frac{\omega}{2\lambda}\,,\:-i\,\frac{\omega-\omega'}{2\lambda})
\,,\\
&&
-\,\tilde{\beta }_{\omega'\omega}^{R } =  \beta_{\omega'\omega}^{R} =
-\,\frac{1}{2\lambda}\,\sqrt{\frac{\omega}{\omega'}}\,
(\frac{M}{2\lambda})^{-i\frac{\omega+\omega'}{2\lambda}}\,
B(-i\,\frac{\omega}{2\lambda}\,,\:i\,\frac{\omega+\omega'}{2\lambda})
 \,, \\
&&
\alpha_{\omega'\omega}^{L} =
-\,\frac{\sinh \frac{\pi(\omega+\omega')}{2\lambda}}{\sinh\frac{\pi\omega'}
{2\lambda}}\,
\beta_{\omega'\omega}^{R} \,, \hspace{0.5cm}
\beta_{\omega'\omega}^{L} =
\frac{\sinh \frac{\pi(\omega-\omega')}{2\lambda}}{\sinh\frac{\pi\omega'}
{2\lambda}}\,
\alpha_{\omega'\omega}^{R} \,,
\end{eqnarray}
where
\( \:
B(x,y) = \frac{\Gamma(x)\Gamma(y)}{\Gamma(x+y)}
\: \)
is the beta function and asymptotically
\begin{equation}
B(x,y) \rightarrow \Gamma(x)\,y^{-x} \,,
\end{equation}
as
\( \:
|y| \rightarrow \infty
\: \).
From these a symmetry relation follows;
\begin{eqnarray}
\beta_{\omega'\omega}^{R} =
\frac{\sinh\frac{\omega'}{2\lambda}}{\sinh\frac{\omega}{2\lambda}}\,
\beta_{\omega\omega'}^{R} \,, \hspace{0.5cm}
\alpha_{\omega'\omega}^{R} =
\frac{\sinh\frac{\omega'}{2\lambda}}{\sinh\frac{\omega}{2\lambda}}\,
\alpha_{\omega\omega'}^{R *} \,.
\end{eqnarray}

\vspace{0.5cm}
\addtocounter{chapter}{1}
\setcounter{equation}{0}
\section*{
C.
Correlations among right-movers}

\vspace{0.5cm}
\hspace*{0.5cm}
The right-moving field with the reflection boundary condition is defined by
\begin{eqnarray}
f^{R}(\sigma)
= \frac{1}{2\pi}\, \int_{0}^{\infty}\,d\omega\,
(f_{\omega}^{R}a_{\omega}^{R}+f_{\omega}^{R *}a_{\omega}^{R\dag}) \,,
\end{eqnarray}
where the right moving out-mode function is
\begin{eqnarray}
f^{R}_{\omega}
&=&
\frac{1}{\sqrt{2\omega}}\, [\:
-\,e^{-i\omega\sigma^{-}}+e^{-i\omega q(\sigma^{+})}\:
\theta(x_{H} - \sigma^{+})\:] \, \nonumber \\
&=&
\frac{1}{\sqrt{2\omega}}\, [\: -\,e^{-i\omega\sigma^{-}}
+ (e^{-2\lambda\sigma^{+}}-\frac{M}{2\lambda})^{i\frac{\omega}{2\lambda}}\:
\theta(x_{H} - \sigma^{+})\:] \,.
\end{eqnarray}
By noting that
\( \:
f_{\omega }^{R *} = f_{-\omega }^{R} \,,
\: \)
the two point correlator made of these right movers is calculated as
\begin{eqnarray}
&&
\langle\, f^{R}(\sigma)f^{R}(\sigma')\, \rangle\:_{\rm{in}} =
\int_{0}^{\infty}\, \frac{d\omega}{2\pi}\,
\int_{0}^{\infty}\, \frac{d\omega'}{2\pi}\,
\int_{0}^{\infty}\, \frac{d\omega''}{2\pi}\,
\nonumber \\
&&
\left(
f_{\omega}^{R*}\,\langle\, f_{\omega''}^{*} || f_{\omega}^{R}\, \rangle
- f_{-\omega}^{R*}\,\langle\, f_{\omega''}^{*} || f_{-\omega}^{R}\, \rangle
\right)
\left( f_{\omega'}^{R}\,\langle\, f_{\omega'}^{R} ||
f_{\omega''}^{*}\, \rangle - f_{-\omega'}^{R}\,\langle\,
f_{-\omega'}^{R} || f_{\omega''}^{*}\, \rangle \right)
\,, \label{rcorrelator}
\end{eqnarray}
with the Klein-Gordon inner product given by
\begin{equation}
\langle\, f_{1} || f_{2}\, \rangle =
i \int\, d\sigma^{\mu}\,(f_{1}^{*}\partial_{\mu}f_{2} -
f_{2}\partial_{\mu}f_{1}^{*}) \,.
\end{equation}
By noting that the original integrand is
\begin{equation}
\left(
f_{\omega}^{R*}\,\tilde{\beta}_{\omega''\omega}^{R*} +
f_{\omega}^{R}\,\tilde{\alpha}_{\omega''\omega}^{R}
 \right)
\left(
\tilde{\beta}_{\omega''\omega'}^{R}\,f_{\omega'}^{R} +
\tilde{\alpha}_{\omega''\omega'}^{R*}\,f_{\omega'}^{R*}
 \right) \,,
\end{equation}
and using the completeness relation Eq.\ref{completeness},
one may write the correlation
as in Eq.\ref{correlationg} of the text.

Explicitly, the mode sum is performed as
\begin{eqnarray}
&&
\int_{0}^{\infty}\, \frac{d\omega}{2\pi}\,
[\: f_{\omega}^{R*}(\sigma) f_{\omega}^{R}(\sigma_{1})
- f_{-\omega}^{R*}(\sigma) f_{-\omega}^{R}(\sigma_{1}) \:] =
\nonumber \\
&&
\hspace*{-0.4cm}
\frac{1}{4\pi}\,\ln\,
[\,\frac{\sigma^{-} -\sigma_{1}^{-} -i\epsilon}{\sigma_{1}^{-}
-\sigma^{-} -i\epsilon}\,]\,
[\,\frac{q(\sigma^{+}) - q(\sigma_{1}^{+}) -i\epsilon}
{q(\sigma_{1}^{+}) - q(\sigma^{+}) -i\epsilon}\,]\,
[\,\frac{q(\sigma_{1}^{+}) - \sigma^{-} -i\epsilon}
{\sigma^{-} - q(\sigma_{1}^{+}) -i\epsilon}\,]\,
[\,\frac{\sigma_{1}^{-} - q(\sigma^{+}) -i\epsilon}
{q(\sigma^{+}) - \sigma_{1}^{-} -i\epsilon}\,]
\nonumber \\
&&
=
\frac{i}{4}\,[\,
\epsilon(\sigma^{-} - \sigma_{1}^{-})
+\epsilon(\,q(\sigma^{+}) - q(\sigma_{1}^{+})\,)
-\epsilon(\sigma^{-} - q(\sigma_{1}^{+})\,)
-\epsilon(\,q(\sigma^{+}) - \sigma_{1}^{-})\,] \,, \nonumber \\
\end{eqnarray}
with $\epsilon (x)$ the signature function :
\( \:
\epsilon (x) = 1 \; \mbox{for} \: x>0 \,, \hspace{0.2cm} \mbox{and} \:
= -1 \; \mbox{for} \: x<0
\: \).
The inverted function $q(\sigma^{+})$ is not defined for \( \:
\sigma^{+} > x_{H} \,,
\: \) thus is to be discarded in this region.
Both infrared and ultraviolet divergences are cancelled out in this formula.
Ignoring irrelevant infinities and writing the correlator
only in the region of \( \:
\sigma^{+} > x_{H}
\: \),
\begin{eqnarray}
\langle\, f^{R}(\sigma)f^{R}(\sigma')\,\rangle \:_{\rm{in}} =
-\,\frac{1}{8\pi}\, \ln\,\frac{
[\,p(\sigma^{-}) - p(\sigma'^{-}) - i\epsilon\,]^{2}}
{[\,x_{H} - p(\sigma'^{-}) - i\epsilon\,]\:
[\,p(\sigma^{-}) - x_{H} - i\epsilon\,]}
 \,.
\end{eqnarray}
Thus observable correlators are identical to those of the complete field
\begin{equation}
\langle\, \partial_{-}f^{R}(\sigma)\partial_{-}f^{R}(\sigma')\,\rangle
\:_{\rm{in}} =
\langle\, \partial_{-}f(\sigma)\partial_{-}f(\sigma')\,\rangle
\:_{\rm{in}} =
-\,\frac{1}{4\pi}\,\frac{p'(\sigma^{-})\,p'(\sigma'^{-})}
{[\,p(\sigma^{-}) - p(\sigma'^{-}) - i\epsilon\,]^{2}} \,.
\end{equation}
This identity may be proved valid also for an extended region of \( \:
\sigma^{+} < x_{H}
\: \).
This should be expected, because \( \:
\partial_{-}f = \partial_{-}f^{R}
\: \)
as an operator identity.

\vspace{0.5cm}
\addtocounter{chapter}{1}
\setcounter{equation}{0}
\section*{
D.
Definition and simple properties of Schwartzian derivative}

\vspace{0.5cm}
\hspace*{0.5cm}
The Schwartzian derivative is defined by
\begin{eqnarray}
&&
\{ f \,,\: x\}_{S} = \frac{\partial_{x}^{3}f }{\partial_{x}f} -
\frac{3}{2}\,(\frac{\partial_{x}^{2}f}{\partial_{x}f})^{2}\,,
\end{eqnarray}
where $f$ is regarded as a function of $x$ : $f(x)\:$, with
the notation
\( \:
\partial_{x} = \frac{\partial }{\partial x} \,.
\: \)
Some simple properties of the Schwartzian derivative
may be derived by a straightforward computation;
\begin{eqnarray}
&&
\{ x \,,\: f\}_{S} = -\,(\frac{dx}{df})^{2}\, \{ f \,,\: x\}_{S} \,, \\
&&
\{ af \,,\: x\}_{S} = \{ f \,,\: x\}_{S} \,, \\
&&
\{ f \,,\: ax\}_{S} = \frac{1}{a^{2}}\,\{ f \,,\: x\}_{S} \,, \\
&&
\{ f(y) \,,\: x\}_{S} = (\frac{\partial y}{\partial x})^{2}\,\{ f \,,\: y\}_{S}
+ \{ y \,,\:x\}_{S} \,, \\
&&
\{ f \,,\: x(z)\}_{S} = (\frac{\partial z}{\partial x})^{2}\,[\:
\{ f \,,\: z\}_{S} - \{ x \,,\: z\}_{S} \:] \,,
\end{eqnarray}
where $a$ is a constant.

\def\thebibliography#1{\list
 {[\arabic{enumi}]}{\settowidth\labelwidth{[#1]}\leftmargin\labelwidth
 \advance\leftmargin\labelsep
 \usecounter{enumi}}
 \def\newblock{\hskip .11em plus .33em minus .07em}
 \sloppy\clubpenalty4000\widowpenalty4000
 \sfcode`\.=1000\relax}
\let\endthebibliography=\endlist
\def\AP{{\sl Ann.\ Phys.\ {\rm(}N.Y.{\rm)} }}
\def\CMP{{\sl Commun.\ Math.\ Phys.\ }}
\def\FP{{\sl Fortsch.\ Phys.\ }}
\def\GRG{{\sl Gen.\ Rel.\ Grav.\ }}
\def\JMP{{\sl J.\ Math.\ Phys.\ }}
\def\JPSJ{{\sl J.\ Phys.\ Soc.\ Jpn.\ }}
\def\LNC{{\sl Lett.\ Nuovo Cim.\ }}
\def\LNCI{{\sl Lett.\ Nuovo Cim.\ }(Ser.~I), }
\def\MPL{{\sl Mod.\ Phys.\ Lett.\ }}
\def\MPLA{{\sl Mod.\ Phys.\ Lett.\ A }}
\def\NC{{\sl Nuovo Cimento }}
\def\NCA{{\sl Nuovo Cimento A }}
\def\NP{{\sl Nucl.\ Phys.\ }}
\def\NPB{{\sl Nucl.\ Phys.\ B }}
\def\PL{{\sl Phys.\ Lett.\ }}
\def\PLA{{\sl Phys.\ Lett.\ A }}
\def\PLB{{\sl Phys.\ Lett.\ B }}
\def\PRep{{\sl Phys.\ Rep.\ }} \let\PREP=\PRep
\def\PR{{\sl Phys.\ Rev. }}
\def\PRA{{\sl Phys.\ Rev.\ A }}
\def\PRD{{\sl Phys.\ Rev.\ D }}
\def\PRL{{\sl Phys.\ Rev.\ Lett.\ }}
\def\PS{{\sl Physica Scripta }}
\def\PTP{{\sl Prog.\ Theor.\ Phys.\ }}
\def\PTPS{{\sl Prog.\ Theor.\ Phys.\ Suppl.\ }}
\def\ZPC{{\sl Z.\ Phys.\ C\ }}

\newpage

\begin{center}
\begin{Large}
{\bf Figure Caption}
\end{Large}
\end{center}

\vspace{1cm}
\begin{large}
\hspace*{-1cm}
Fig. 1
\end{large}

Schematic diagram that exhibits the null lines for light rays
reflected by the moving mirror in the mirror spacetime.
$\sigma^{\pm }$ here is the light cone coordinate.
The hatched region to the left of the mirror is bounded by the mirror
trajectory
\( \:
\sigma^{+} = p(\sigma^{-}) \,.
\: \)
$I^{-}$ is the past null infinity, while $I_{R}^{+}$ and $I_{L}^{+}$
are the right and the left future null infinities.

\vspace{1cm}
\begin{large}
\hspace*{-1cm}
Fig. 2
\end{large}

Energy momentum tensor
\( \:
\langle \, T_{--} \, \rangle
\: \)
observable at $I_{R}^{+}$ in unit of $4\lambda^{2}$,
plotted against the light cone time $\sigma^{-}$
in unit of $2\lambda $.
Shown here are results of numerical computation for various models:
the semiclassical case without the back reaction denoted by classical
limit, the cases incorporating the back reaction with the parameter
\( \:
\gamma = 0 \,, \; -\,0.4 \,, \; \mbox{and} -\,0.5 \,,
\: \)
assuming a relatively weak back reaction of large $mM$.
The semiclassical value of the asymptotic flux is indistinguishable
in this figure, but actually different from the case of $\gamma = -\,0.5$.
Chosen parameters are
\( \:
\frac{6\pi m}{\lambda } = 1 \,, \hspace{0.2cm} \mbox{and} \hspace{0.2cm}
\frac{M}{2\lambda } = 10^{8} \,.
\: \)

\vspace{1cm}
\begin{large}
\hspace*{-1cm}
Fig. 3
\end{large}

Energy momentum tensor
\( \:
\langle \, T_{--} \, \rangle
\: \)
observable at $I_{R}^{+}$
with a much stronger back reaction  of small $mM$ than in Fig. 2.
Chosen parameters are
\( \:
\frac{6\pi m}{\lambda } = 1 \,, \hspace{0.2cm} \mbox{and} \hspace{0.2cm}
\frac{M}{2\lambda } = 1 \,.
\: \)

\vspace{1cm}
\begin{large}
\hspace*{-1cm}
Fig. 4
\end{large}

Mirror trajectory given in terms of the light cone variables $z^{\pm }$
for the same set of parameters as in Fig. 3.

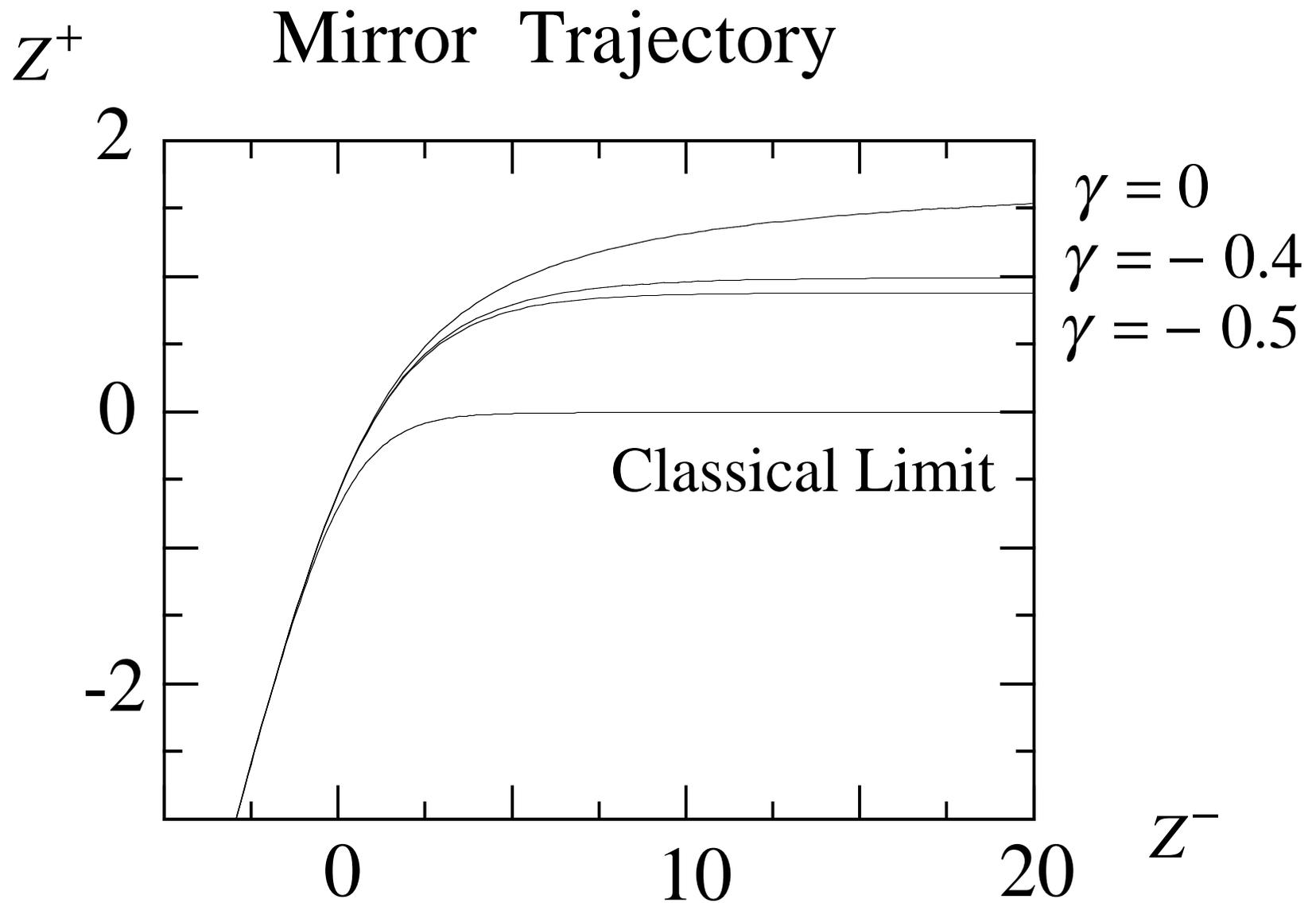
\end{document}